\documentclass[aps,prx,twocolumn,superscriptaddress,amssymb,amsfonts,nofootinbib]{revtex4-2}

\usepackage{color}
\usepackage{verbatim}
\usepackage{amsmath}
\usepackage{amssymb}
\usepackage{dsfont}
\usepackage{graphicx}
\usepackage{esint}
\usepackage{esvect}
\usepackage{braket}
\usepackage{amsmath,bm}
\usepackage{graphicx}
\usepackage{subfigure}
\usepackage{url}
\usepackage{bbm}
\usepackage[dvipsnames]{xcolor}
\usepackage[
    colorlinks,
    linkcolor={blue!80!black},
    citecolor={blue!80!black},
    urlcolor={blue!80!black}
]{hyperref}

\makeatletter
\@ifundefined{textcolor}{}
{%
 \definecolor{BLACK}{gray}{0}
 \definecolor{WHITE}{gray}{1}
 \definecolor{RED}{rgb}{1,0,0}
 \definecolor{GREEN}{rgb}{0,1,0}
 \definecolor{BLUE}{rgb}{0,0,1}
 \definecolor{CYAN}{cmyk}{1,0,0,0}
 \definecolor{MAGENTA}{cmyk}{0,1,0,0}
 \definecolor{YELLOW}{cmyk}{0,0,1,0}
}

\newcommand{\bR}{\mathbf{R}}
\newcommand{\bk}{\mathbf{k}}
\newcommand{\btau}{\boldsymbol{\tau}}
\newcommand{\br}{\mathbf{r}}
\newcommand{\bx}{\mathbf{x}}
\newcommand{\bq}{\mathbf{q}}

\newcommand{\bh}{\mathbf{h}}

\newcommand{\bp}{\mathbf{p}}

\newcommand{\Tr}{\text{Tr}}

\makeatother
\begin{document}

\title{Orbital Wigner functions and quantum transport in multiband systems}

\author{Johannes Mitscherling}
\thanks{These authors contributed equally.}
\affiliation{Institute for Theoretical Physics and Astrophysics, University of W\"urzburg, and W\"urzburg-Dresden Cluster of Excellence ctd.qmat, D-97074 W\"urzburg, Germany}
\affiliation{Max Planck Institute for the Physics of Complex Systems, N\"othnitzer Str. 38, 01187 Dresden, Germany}
\affiliation{Department of Physics, University of California, Berkeley, California 94720, USA}
\author{Dan S. Borgnia}
\thanks{These authors contributed equally.} 
\affiliation{Department of Physics, University of California, Berkeley, California 94720, USA}
\author{SuryaNeil Ahuja}
\affiliation{Department of Physics, University of California, Berkeley, California 94720, USA}
\author{Joel E. Moore}
\affiliation{Department of Physics, University of California, Berkeley, California 94720, USA}
\affiliation{Materials Sciences Division, Lawrence Berkeley National Laboratory, Berkeley, California 94720, USA}
\author{Vir B. Bulchandani}
\affiliation{Department of
Physics and Astronomy, Rice University, 6100 Main Street
Houston, Texas 77005, USA}
\begin{abstract}
Traditional theories of electron transport in crystals are based on the Boltzmann equation and do not capture physics arising from quantum coherence. We introduce a transport formalism based on orbital Wigner functions, which accurately captures quantum coherent physics in multiband fermionic systems. We illustrate the power of this approach compared with traditional semiclassical transport theory by testing it numerically against microscopic simulations of one-dimensional, noninteracting, two-band systems---the simplest systems capable of exhibiting interorbital coherence. We show that orbital Wigner functions accurately capture strongly nonequilibrium features of electron dynamics that lie beyond conventional Boltzmann theory, such as the ballistic transport of a relative phase between microscopic orbitals and topological Thouless pumping of charge both at nonzero temperature and away from the adiabatic limit. Our approach is motivated in part by modern ultracold atom experiments that can prepare and measure far-from-equilibrium charge transport and phase coherence in multiband fermionic systems, calling for correspondingly precise theories of transport. The quantitative accuracy exhibited by our approach, together with its capacity to capture nontrivial physics even at the ballistic scale, establishes orbital Wigner functions as an ideal starting point for developing a fully systematic theory of transport in crystals.
\end{abstract}
\maketitle

\section{Introduction}
\subsection{Background and motivation}
The semiclassical theory of electron transport in crystals has changed little since its development by Drude, Sommerfeld and others in the early part of the previous century~\cite{ashcroft1976solid}. Perhaps its most significant revision was the inclusion of Berry-curvature effects such as the Karplus-Luttinger anomalous velocity~\cite{Chang_1995,Chang_1996,sundaram1999wave}, many decades after the anomalous velocity was discovered~\cite{karplus1954hall}, culminating in a successful theory of the anomalous Hall effect~\cite{jungwirth2002anomalous,Sinitsyn_2007,nagaosa2010anomalous}. It seems remarkable in hindsight that such an essential feature of electron transport was missed for so long. One possible explanation for this delay is the well-known lack of a fully systematic derivation of the semiclassical equations of motion for electron transport, despite their traditional justification from wavepackets of Bloch electrons~\cite{ashcroft1976solid}. Nevertheless, a suitably careful analysis of such Bloch-electron wavepackets turned out to be sufficient to capture Berry-curvature effects~\cite{Chang_1995,Chang_1996,sundaram1999wave}, and the resulting equations of motion form the bedrock of the modern theory of transport.

Despite correctly accounting for Berry-curvature effects, semiclassical transport theory is less reliable above zero temperature and in the presence of multiple bands. Quantum coherences between different single-particle bands become possible, whose accurate description requires a matrix-valued transport equation rather than one Boltzmann equation per band as arises when quantum coherences are neglected~\cite{ashcroft1976solid}. Systematic approaches to quantum transport in multiband systems range from analyses using the non-Abelian Berry curvature in the special case of degenerate or near-degenerate bands~\cite{culcer2005coherent,shindou2005noncommutative,Stedman2020} to more general Wigner-function based approaches~\cite{demeio2002wigner,unlu2004multi,Culcer2006,Culcer2009,Morandi2009,Wong2011,Wickles2013,Morawetz2015,Iafrate2017,Sekine2017,Cepellotti2021,Bhalla2021,Konig2021,Valet2023} that model time evolution of the full single-particle density matrix and not just its diagonal elements as tradition dictates.

Such Wigner-function-based approaches to quasiparticle transport have a long history in physics from the pioneering work of Irving and Zwanzig~\cite{irving1951statistical}. However, they have only been applied to electrons in crystal lattices more recently, with two notable strands of work in this area. One popular approach in solid-state physics (see Refs.~\cite{demeio2002wigner,unlu2004multi,Culcer2006,Culcer2009,Morandi2009,Wong2011,Wickles2013,Morawetz2015,Iafrate2017,Sekine2017,Cepellotti2021,Bhalla2021,Konig2021,Valet2023} and references therein) consists of deriving dynamical equations for Wigner functions of Bloch electrons. While this method has proved effective for capturing a wide variety of multiband effects, such derivations are not usually performed with an accurate characterization of spatially inhomogeneous or quantum coherent dynamics in mind. This reflects the complex physics of real materials. (i) An accurate theoretical description of the position-space charge density can be extremely challenging, even at the level of unambiguously interpreting scanning tunneling microscopy (STM) data~\cite{holbrook2024realspaceimagingbandtopology}, and (ii) quantities that are typically measured in transport experiments on solid-state systems are averages over momentum space near thermal equilibrium, so that approximation schemes yielding equivalent predictions for equilibrium transport can predict significantly different dynamics from one another in spatially inhomogeneous and quantum coherent regimes. Nevertheless, some of these formalisms explicitly account for the distribution of charge in position space in a controlled and gauge-invariant fashion~\cite{Culcer2006,Wickles2013}.

A somewhat different perspective on deriving transport from Wigner functions has arisen in the setting of single-particle lattice models, for which a definition of a Wigner function with the correct position-space marginals was obtained by Hinarejos {\it et al.}~\cite{hinarejos2012wigner}. While this development appears to have escaped notice in most literature on Bloch-electron transport, it was pursued further in the setting of one-dimensional (1D) systems, where it now forms the basis for a quantitatively accurate and thoroughly tested theory of transport~\cite{fagotti2017higher,Bastianello_2018,Fagotti_2020,Coppola_2023,essler2023short}. However, the latter tends to focus on single-band systems, for which nontrivial physical effects from interband quantum coherence cannot arise.

In this paper, we seek to bridge this gulf between the richness of multiband transport in arbitrary crystal lattices and the rigor of recent approaches to transport in 1D. We achieve this by proposing {\it orbital Wigner functions} as an alternative to previous Bloch-electron-based constructions of Wigner functions in crystal lattices. Such orbital Wigner functions have their position-space marginals located at Wannier centers, which is a natural generalization of the construction by Hinarejos {\it et al.}~\cite{hinarejos2012wigner,hinarejos2015wigner} to arbitrary crystal lattices. We view this property as a prerequisite for obtaining predictions for the space-time electron density at the same level of quantitative accuracy as the underlying tight-binding description. This implies that our method is only as accurate as the {\it ab initio} model that underlies the tight-binding Hamiltonian in question, a limitation that it shares with the vast majority of transport theories in solid-state physics. On the other hand, if the tight-binding description of a given system is accurate, our method introduces no further approximations beyond the ballistic scaling limit~\cite{spohn2012large} and should therefore accurately characterize large-scale dynamics in both spatially inhomogeneous and quantum coherent regimes. Thus we expect the results of this paper to be most useful for atomic, molecular and optical (AMO) systems, for which particles reside on a tunable optical lattice, with a tunable number of particles per site and tunable interparticle interactions mediated by Feshbach resonances, enabling the direct observation of far-from-equilibrium position-space dynamics in agreement with noninteracting tight-binding predictions (see e.g., Refs.~\cite{nakajima2016topological,fujiwara2019transport}). Given these experimental capabilities and in the interests of thoroughly benchmarking our proposed approach in these challenging quantum coherent regimes, we will restrict our attention to noninteracting systems in this paper and return to the topic of interactions in the conclusion.

Orbital Wigner functions appeared previously as an intermediate step in a recent derivation of a transport equation for electrons (see supplementary data of Ref.~\cite{Cepellotti2021}), although the main result reported in that work, which was expressed in terms of Bloch electrons, does not appear to be consistent with the results reported here. A formalism equivalent to ours was also proposed recently for phonons~\cite{Simoncelli2022}. However, the authors of the latter paper benchmarked the proposed equation by introducing collision terms for comparison with experimental data, thereby introducing physics beyond the scope of this work. We instead benchmark our proposed transport equation against microscopic numerical simulations of the simplest non-trivial case, namely 1D, noninteracting, two-band systems.

Even in this apparently simple setting, which involves only streaming rather than collision terms in the Boltzmann language,  we identify physical effects that fall beyond the purview of textbook transport theory, such as time evolution from locally thermal states perturbed by a coherent interorbital phase or Thouless-type~\cite{thouless1983quantization} charge pumping that persists both far from adiabaticity and at nonzero temperature. We nevertheless show in all such cases that the resulting charge and current dynamics is captured to within numerical accuracy by our proposed ballistic-scale transport equation. At the same time, we present numerical evidence and a general analytical argument that our proposed equation recovers standard Boltzmann-theory predictions in local thermal equilibrium and therefore does not contradict the textbook understanding of electron transport~\cite{ashcroft1976solid}. We note that the stringent benchmarking procedure that we follow in this work is applicable in principle to any other proposed transport equation for electrons in crystal lattices and may thus be of independent interest.

The reader might wonder at our emphasis on quantitative accuracy, given that the theory of transport in crystals is often viewed as a coarse-grained and largely qualitative tool. Our motivations are twofold. First, contemporary AMO experiments are capable of imaging spacetime profiles of conserved charge densities in mesoscopic systems with sufficient accuracy to discern their lineshapes \cite{nakajima2016topological,fujiwara2019transport,Brown_2019,Schemmer19,Bakr2,Gross_2021,Malvania_2021,Wei_2022,Brandstetter_2025} and therefore demand similarly quantitatively accurate theoretical predictions. Second, all the difficulties of transport theory are usually held to lie in its collision terms. 
The fact that multiband systems exhibit nontrivial physics at the level of their streaming terms indicates that more care is needed to capture quantum coherent effects, even at the ballistic or Euler scale~\cite{spohn2012large} where dissipative corrections can be ignored. 

\subsection{Guide to results}

We introduce a transport theory based on orbital Wigner functions for generic tight-binding Hamiltonians on $d$-dimensional crystal lattices in Sec.~\ref{sec:owf}. The matrix-valued orbital Wigner function $w(\bx,\bk)$ defined in Eq.~\eqref{eqn:OrbitalWignerFunction} allows us to capture local densities of conserved operators, such as the local charge density and local charge-current density, even when the initial state is far from equilibrium and quantum coherent. Starting from the exact Heisenberg equations of motion for fermion bilinears, we derive a coupled system of linear partial differential equations (PDEs) that captures the ballistic-scale dynamics of the orbital Wigner function, see Eq.~\eqref{eqn:EoM}. Our framework generalizes straightforwardly to time-dependent tight-binding Hamiltonians and external electrostatic potentials, see Eqs.~\eqref{eqn:tdepWignerEoM} and \eqref{eqn:FieldEoM} respectively.

We check the validity of our proposed transport equation against numerical free-fermion simulations in Secs.~\ref{sec:ComparsionSemiClassic} and \ref{sec:timedependence} for some representative Hamiltonians, parameter regimes, and initial states that lie beyond the scope of semiclassical transport theory. For example, we demonstrate that our transport equation accurately captures dynamics resulting from initial states with spatially inhomogeneous single-particle coherence (Fig.~\ref{Fig:PhaseCohCompn}), coherently driven Hamiltonians even for fast driving frequencies (Fig.~\ref{fig:loceqtdep}), and the charge dynamics in topological Thouless pumps over multiple pumping cycles and far from the idealized adiabatic limit (Figs.~\ref{fig:ThoulessCurr} and \ref{fig:ThoulessTest}), despite reducing to conventional Boltzmann theory in local thermal equilibrium as expected (Fig.~\ref{Fig:BoltzmannComp}). We also model stroboscopically driven Su-Schrieffer-Heeger (SSH) chains whose Bloch bands are always instantaneously flat but nevertheless allow for far-from-equilibrium charge transport that is captured by our formalism (Fig.~\ref{fig:FlatBands}). Our results thus provide a unified starting point for making quantitative predictions about multiband dynamics and transport in state-of-the-art analog quantum simulators and in advanced experiments on quantum materials, in far-from-equilibrium regimes that lie well beyond standard theoretical simplifications.

We close in Sec.~\ref{sec:discussion} with a discussion of the possibilities opened up by our approach, which range from a systematic treatment of Berry curvature and further quantum geometric effects for multiple bands away from zero temperature to natural extensions of our transport equation to include disorder or interactions, for example, through collision terms that correctly account for quantum coherence in multiband systems.

\section{Orbital Wigner functions}
\label{sec:owf}
The main results of this section are (i) an equation of motion for Wigner functions defined in the orbital or Wannier basis given in Eq.~\eqref{eqn:EoM} and (ii) its generalization to time-dependent Hamiltonians given in Eq.~\eqref{eqn:tdepWignerEoM}. These equations form the basis of the subsequent analysis in Secs.~\ref{sec:ComparsionSemiClassic} and \ref{sec:timedependence}. The straightforward generalization to static electric 
fields is presented in Appendix~\ref{appendix:extfield}, where we provide the ballistic-scale equation of motion in the presence of an external electrostatic potential in Eq.~\eqref{eqn:FieldEoM}.
\subsection{Hamiltonian conventions}
\label{sec:tbc}
We begin by fixing conventions for the fermionic tight-binding Hamiltonians that we will consider in this work and their associated densities of charge, charge current and energy. 
\subsubsection{Tight-binding Hamiltonians}

Let the index $j \in \Lambda$ label the sites $\mathbf{R}_j \in \mathbb{R}^d$ of a $d$-dimensional Bravais lattice and let $\btau_{\alpha}$ denote the positions of additional degrees of freedom relative to $\mathbf{R}_j$ within each unit cell, with $\alpha=1,2,\ldots,M$. The latter may reflect spin states, atomic orbitals, or other basis sites, all of which we refer to as {\it orbitals}  for simplicity. Throughout the paper we set $\hbar=1$ for convenience.

It will be useful to write $\mathbf{R}_{j\alpha} = \mathbf{R}_j + \btau_{\alpha}$ for the absolute position of each orbital. A generic tight-binding Hamiltonian $\hat{H}_0$ associated with this crystal structure can be written as
\begin{align}
    \hat{H}_0 = -\sum_{j,j'\in \Lambda}\sum_{\alpha,\beta=1}^M t^{\alpha\beta}_{jj'}\,\, \hat c^\dagger_{j\alpha}\hat c^{}_{j'\beta} \, .
    \label{eqn:Hamiltonian}
\end{align}
We will assume lattice translation invariance, so that the hopping amplitudes depend only on the relative positions $t^{\alpha\beta}_{jj'} = t_{\alpha\beta}(\bR_j-\bR_{j'})$. We also assume Hermiticity of the Hamiltonian, $t^{\alpha\beta}_{jj'} = (t^{\beta\alpha}_{j'j})^*$. For simplicity, we consider only finite-range hopping. Note that this Hamiltonian conserves the total charge $\hat{N} = \sum_{j\in\Lambda} \sum_{\alpha=1}^M \hat{c}^\dagger_{j\alpha} \hat{c}^{}_{j\alpha}$. Equation~\eqref{eqn:Hamiltonian} represents the most general charge-conserving and noninteracting Hamiltonian for fermions arranged in a crystal lattice and makes no additional assumptions about the underlying Wannier states that give rise to the tight-binding parameters $t^{\alpha\beta}_{jj'}$.

However, at various points in this paper, we will make the additional (albeit standard) assumption that the position of a 
given Wannier orbital is well approximated by the position of its Wannier center. We make this assumption implicitly in Eq.~\eqref{eq:defJ} when defining the center-of-charge operator and its associated charge-current operator and again in Eq.~\eqref{eq:defV} when considering dynamics in external electrostatic fields. In local thermal equilibrium, these simplifying approximations should not affect our predictions at linear order in applied electric fields, but nonlinear optical responses are particularly sensitive to off-diagonal contributions to the center-of-charge or position operator~\cite{Pedersen, Sandu, Iba_ez_Azpiroz_2022}. For a systematic treatment of such off-diagonal contributions within our formalism, we refer to Appendix \ref{appendix:offdiag}.

\subsubsection{Orbital vs Bloch electrons}

We define momentum space fermions by
\begin{equation}
    \hat{c}_{\alpha}(\bk) = \sum_{j \in \Lambda} e^{-i\mathbf{k}\cdot \mathbf{R}_{j\alpha}} \hat{c}_{j\alpha}
\end{equation}
(see Appendix  \ref{appendix:FourierConventions} for a summary of our Fourier transform conventions), in terms of which
\begin{equation}
    \hat{H}_0 = \int_{\mathrm{BZ} }\frac{d\mathbf{k}}{V_{\mathrm{BZ}}} \sum_{\alpha,\beta=1}^M h^{}_{\alpha \beta}(\mathbf{k}) \,\hat{c}_{\alpha}^\dagger(\bk)\,\hat{c}^{}_{\beta}(\bk)\,,
\end{equation}
where $\int_{\mathrm{BZ}}$ denotes integration over the first Brillouin zone (BZ) with volume $V_{\mathrm{BZ}}$, and the Bloch Hamiltonian $h_{\alpha \beta}(\mathbf{k}) = -\sum_{j\in\Lambda} t_{\alpha \beta}(\bR_j) e^{-i\mathbf{k}\cdot (\bR_j +\btau_{\alpha}-\btau_\beta)}$. For each $\mathbf{k}$, there exists an $M$-by-$M$ unitary matrix $U(\mathbf{k})$ diagonalizing the matrix $h(\mathbf{k})$ such that
\begin{equation}
    \label{eq:changeofbasis}
    \sum_{\alpha,\beta=1}^M U^{\dagger}_{m \alpha}(\mathbf{k}) \,h_{\alpha \beta} (\mathbf{k})\,U_{\beta n}(\mathbf{k}) = E_{n}(\mathbf{k})\, \delta_{mn}\,,
\end{equation}
which defines $M$ energy bands $\{E_n(\mathbf{k})\}_{n=1}^M$ and Bloch electron creation operators $\hat{d}^\dagger_{n}(\mathbf{k}) := \sum_{\alpha=1}^M U_{\alpha n}(\bk) \hat{c}^\dagger_{\alpha}(\bk)$, in terms of which
\begin{equation}
    \hat{H}_0 = \int_{\mathrm{BZ}} \frac{d\mathbf{k}}{V_{\mathrm{BZ}}} \sum_{n=1}^M E^{}_n(\mathbf{k}) \,\hat{d}^\dagger_n(\mathbf{k}) \,\hat{d}^{}_n(\mathbf{k}) \,.
\end{equation}
We will refer to the excitations created by $\hat{c}_a^\dagger$ and $\hat{d}_n^\dagger$ as {\it orbital electrons} and {\it Bloch electrons} respectively.

\subsubsection{Operator densities of conserved charges}

It will be useful to define local Hamiltonian and charge density operators
\begin{align}
    \label{eqn:localHamiltonian}
    &\hat{h}_j = -\sum_{j'\in \Lambda}\sum_{\alpha,\beta=1}^M t^{\alpha\beta}_{jj'}\,\, \hat c^\dagger_{j\alpha}\hat c^{}_{j'\beta} \, , 
\end{align}
and
\begin{align}
    \label{eqn:localCharge}
    &\hat{n}_j = \sum_{\alpha=1}^M \hat{c}_{j\alpha}^\dagger \hat{c}^{}_{j\alpha} \, ,
\end{align}
respectively, which satisfy $\hat{H}_0 = \sum_{j \in \Lambda} \hat{h}_j$ and $\hat{N} = \sum_{j\in \Lambda} \hat{n}_j$ respectively. 

We will also have reason to consider the charge-current operator, which is most simply defined as the rate of change of the center-of-charge operator $\hat{\mathbf{X}} = \sum_{j\in\Lambda} \sum_{\alpha=1}^M \mathbf{R}_{j\alpha}\hat{n}_{j\alpha}$, yielding the general expression
\begin{equation}
    \label{eq:defJ}
    \hat{\mathbf{J}} = \frac{d}{dt} \hat{\mathbf{X}} = \int_{\mathrm{BZ}} \frac{d\bk}{V_{\mathrm{BZ}}} \sum_{\alpha,\beta=1}^M \partial^{}_{\bk} h^{}_{\alpha \beta}(\bk) \,\hat{c}^\dagger_{\alpha}(\bk) \,\hat{c}^{}_{\beta}(\bk)\,.
\end{equation}
For the examples of interest in this paper, $\hat{\mathbf{J}} = \sum_{j\in\Lambda} \hat{\mathbf{j}}_j$ is a local operator and we will verify explicitly that the current density operators $\hat{\mathbf{j}}_j$ satisfy a microscopic continuity equation that justifies the definition Eq.~\eqref{eq:defJ}. As discussed above, Eq.~\eqref{eq:defJ} neglects possible off-diagonal contributions to the center-of-charge operator~\cite{Pedersen, Sandu, Iba_ez_Azpiroz_2022}, and when these are important, local observables should instead be computed using the formalism of Appendix \ref{appendix:offdiag}.

\subsection{Definition of the orbital Wigner function}
We define the {\it orbital Wigner operator} at position $\bx$ and crystal momentum $\bk$ by
\begin{equation}
    \label{eq:OrbitalWignerOperator}
    \hat{W}_{\alpha \beta}(\mathbf{x},\mathbf{k}) = \int_{\mathrm{BZ}} \frac{d\mathbf{q}}{V_{\mathrm{BZ}}} \, e^{i\mathbf{q}\cdot \bx}\, \hat{c}^\dagger_\alpha(\mathbf{k}-\mathbf{q}/2)\,\hat{c}^{}_\beta(\mathbf{k}+\mathbf{q}/2) \,.
\end{equation}
The momentum and position marginals of this Wigner operator have two important properties that any single-particle Wigner operator ought to satisfy~\cite{hinarejos2012wigner,hinarejos2015wigner}. First, the momentum marginals satisfy the condition~
\begin{equation}
    \label{eqn:sumW}
    \sum_{j \in \Lambda} \hat{W}_{\alpha \beta}(\bR_{j\alpha},\mathbf{k}) = \hat{c}_{\alpha}^\dagger(\mathbf{k}) \,\hat{c}^{}_{\beta}(\mathbf{k}).
\end{equation}
Second, the integral of diagonal elements of the Wigner operator over pseudomomentum $\mathbf{k}$ at fixed $\bx = \mathbf{R}_{j\alpha}$ is given by
\begin{equation}
    \label{eq:posmarg}
    \int_{\mathrm{BZ}} \frac{d\mathbf{k}}{V_{\mathrm{BZ}}} \hat{W}_{\alpha\alpha}(\mathbf{R}_{j\alpha},\mathbf{k}) = \hat{c}_{j\alpha}^\dagger \, \hat{c}^{}_{j\alpha}\,,
\end{equation}
which is the physical charge density  
associated with the Wannier center $\mathbf{R}_{j\alpha}$. The latter property means that an accurate model for the dynamics of $\hat{W}_{\alpha \beta}$ will yield correspondingly accurate predictions for the dynamics of the local charge density. Finally, a (matrix-valued) \emph{orbital Wigner function} $w_{\alpha\beta}(\bx,\bk)$ can be defined from the orbital Wigner operator by fixing an initial density matrix $\hat \rho$, and considering the expectation value
\begin{align}
    w_{\alpha\beta}(\bx,\bk) := \big\langle \hat{W}_{\beta\alpha}(\bx,\bk)\big\rangle\,,
    \label{eqn:OrbitalWignerFunction}
\end{align}
where we use the notation $\langle \hat{O} \rangle = \mathrm{Tr}[\hat{\rho} \hat{O}]$ for expectation values of operators $\hat{O}$ throughout this paper. This definition can be seen as a generalization of the lattice Wigner functions discussed in Refs. \cite{hinarejos2012wigner,hinarejos2015wigner,Bastianello_2018,Coppola_2023,essler2023short} and the block Wigner functions introduced for spin-$1/2$ chains in Refs. \cite{fagotti2017higher,Fagotti_2020} to arbitrary crystal structures in any spatial dimension, see also Ref. \cite{Cepellotti2021}.

The reader might wonder why we are not attempting to use Bloch electrons rather than orbital electrons to construct the Wigner operator, as is standard in all Wigner-function-based treatments of Bloch electron dynamics above one spatial dimension that we are aware of (see Refs.~\cite{demeio2002wigner,unlu2004multi,Culcer2006,Culcer2009,Morandi2009,Wong2011,Wickles2013,Morawetz2015,Iafrate2017,Sekine2017,Cepellotti2021,Bhalla2021,Konig2021,Valet2023} and references therein), even those that use orbital electrons as an intermediate step~\cite{Cepellotti2021}. The reason is that if one uses orbital electrons rather than Bloch electrons in the definition of the Wigner operator, see Eq.~\eqref{eq:OrbitalWignerOperator}, the position marginals of the Wigner operator, see Eq.~\eqref{eq:posmarg}, correspond to Wannier centers in position space, which is particularly convenient for applications to position-space dynamics. The importance of this property for Wigner functions of lattice particles was pointed out in Refs.~\cite{hinarejos2012wigner,hinarejos2015wigner}. Although it is possible to restore this property in certain Bloch-electron based approaches through a momentum-dependent shift of the position variable~\cite{Culcer2006,Wickles2013}, Eq.~\eqref{eq:posmarg} circumvents such difficulties entirely.

To evolve orbital Wigner functions in time, we consider the Heisenberg-picture dynamics of the underlying orbital Wigner operator. For the tight-binding Hamiltonian Eq.~\eqref{eqn:Hamiltonian}, this yields
\begin{align}
    \partial_t\, \hat W_{\alpha\beta}(\bx,\bk,t)=i\big[\hat H_0,\hat W_{\alpha\beta}(\bx,\bk,t)\big]\,,
    \label{eqn:HeisenbergEoM}
\end{align}
which defines the time-dependent orbital Wigner function $w_{\alpha\beta}(\bx,\bk,t) := \langle \hat{W}_{\beta\alpha}(\bx,\bk,t)\rangle$.

\subsection{Dynamics of densities of conserved charges}
\label{sec:dynCS}
We now discuss how to compute the dynamics of charge densities from the orbital Wigner function $w(\bx,\bk,t)$. By Eq.~\eqref{eq:posmarg}, it follows that the local charge density $n(\mathbf{R}_j,t) := \langle \hat{n}_j(t)\rangle$ at a given lattice site can be expressed as
\begin{equation}
    \label{eq:Wignerpredn}
    n(\bR_j,t) = \int_{\mathrm{BZ}} \frac{d\bk}{V_{\mathrm{BZ}}} \Tr\big[w(\bR_j,\bk,t)\big]\,.
\end{equation}
(Note that the trace here is over internal orbital indices rather than the full Hilbert space.)

More generally, for any translation-invariant fermion bilinear $\hat{Q} = \sum_{j\in \Lambda} \hat{q}_j$ with
\begin{equation}
    \hat{Q} = \int_{\mathrm{BZ}} \frac{d\bk}{V_{\mathrm{BZ}}} \sum_{\alpha,\beta=1}^M q^{}_{\alpha \beta}(\bk)\,\hat{c}^\dagger_{\alpha}(\bk)\,\hat{c}^{}_{\beta}(\bk)\,,
\end{equation}
the expectation value of the total charge $\hat{Q}$ is given by
\begin{equation}
    \label{eq:exactQWig}
    \big\langle \hat{Q}(t) \big\rangle = \sum_{j \in \Lambda}  \int_{\mathrm{BZ}} \frac{d\bk}{V_{\mathrm{BZ}}} \sum_{\alpha,\beta=1}^M q_{\alpha \beta}(\bk) \,w^{}_{\beta \alpha}(\bR_j,\bk,t)
\end{equation}
by Eq.~\eqref{eqn:sumW}. However, for bilinears $\hat{Q}$ whose densities $\hat{q}_j$ extend over multiple sites, the global-to-local correspondence is generally less simple than suggested by Eq.~\eqref{eq:Wignerpredn}. For example, the approximation
\begin{equation}
\label{eq:Wignerpredq}
    q(\bR_j,t) \approx \int_{\mathrm{BZ}} \frac{d\bk}{V_{\mathrm{BZ}}} \Tr\big[w(\bR_j,\bk,t)\,q(\bk)\big]
\end{equation}
to the density $q(\mathbf{R}_j,t) := \langle \hat{q}_j(t)\rangle$ generally holds only up to lattice-scale corrections. This is true for the charge-current density studied in this paper, for which Eq.~\eqref{eq:defJ} predicts the approximation
\begin{equation}
\label{eq:Wignerpredj}
    \mathbf{j}(\bR_j,t) \approx \int_{\mathrm{BZ}} \frac{d\bk}{V_{\mathrm{BZ}}} \, \mathrm{Tr}\big[w(\bR_j,\bk,t)\,\partial^{}_\bk h(\bk,t)\big]
\end{equation}
to the local charge-current density $\mathbf{j}(\bR_j,t) = \langle \hat{\mathbf{j}}_j(t) \rangle$.

It is sometimes possible to derive analogs of the relation Eq.~\eqref{eq:Wignerpredn} [and its underlying identity in Eq.~\eqref{eq:posmarg}] for other fermion bilinears, including the charge current. However, this must be done on a model-dependent basis. For the ballistic-scale dynamics and observables of interest in this paper, the lattice-scale corrections to Eq.~\eqref{eq:Wignerpredq} are negligibly small and can safely be ignored. A more general microscopic treatment of $n$ and $\mathbf{j}$, which does not make the perfect tight-binding assumption that Wannier orbitals are perfectly localized at Wannier centers, is given in Appendix \ref{appendix:offdiag}.

\subsection{Derivation of ballistic-scale transport equation} 
We now summarize the derivation of a ballistic-scale transport equation from the Heisenberg equation of motion Eq.~\eqref{eqn:HeisenbergEoM}, following recent derivations of the Boltzmann equation for lattice Wigner functions in one spatial dimension~\cite{hinarejos2015wigner,fagotti2017higher,essler2023short}. To derive a transport equation for $w_{\alpha\beta}$, note that the Heisenberg-picture evolution of momentum-space fermions is given by $i[\hat{H}_0,\hat{c}_{\alpha}(\bk,t)] = -i \sum_{\gamma=1}^M h_{\alpha \gamma}(\bk) \hat{c}_\gamma(\bk,t)$, which yields
\begin{align}
    \partial_t \,&\hat W^{}_{\alpha\beta}(\bx,\bk,t)=i \sum_{\gamma=1}^M \int_{BZ} \frac{d\bq}{V_{\mathrm{BZ}}} \,e^{i\bq\cdot \bx}\, \nonumber \\ \times&\Big(h^{}_{\gamma\alpha}(\bk-\bq/2)\,\, \hat c^\dagger_\gamma(\bk-\bq/2)\,c^{}_\beta(\bk+\,\bq/2) \nonumber \\& -h^{}_{\beta\gamma}(\bk+\bq/2)\,\, \hat c^\dagger_\alpha(\bk-\,\bq/2)\,c^{}_\gamma(\bk+\bq/2)\Big)\,,
\end{align}
where we used Hermiticity of the Bloch Hamiltonian $h_{\alpha\beta}(\mathbf{k})^* = h_{\beta\alpha}(\mathbf{k})$. Using the definition of the Wigner operator to close this system of equations then yields
\begin{align}
     &\partial_t \hat W_{\alpha\beta}(\bx,\bk,t) \nonumber \\ &=-i\sum_{\gamma=1}^M \sum_{\br \in \Lambda}\,\Big(t^{\gamma\alpha}(\br)\,e^{-i\bk\cdot \br_{\gamma\alpha}}\,\hat W_{\gamma\beta}\big(\bx+\br_{\gamma\alpha}/2,\bk,t\big)\nonumber\\&\hspace{14mm}-t^{\beta\gamma}(\br)\,e^{-i\bk\cdot \br_{\beta\gamma}}\,\hat W_{\alpha\gamma}\big(\bx-\br_{\beta\gamma}/2,\bk,t\big)\Big)\,,
     \label{eqn:EoMSpatial}
\end{align}
where $\br_{\alpha\beta} := \br+\btau_\alpha-\btau_\beta$ denotes relative spatial positions within the crystal lattice.

So far, our analysis has been exact. We now Taylor expand about $\bx$ to yield a derivative expansion in the spatial gradient $\partial_{\mathbf{x}}$. A full discussion, including higher-order corrections in $\partial_{\mathbf{x}}$, is given in Appendix~\ref{appendix:HigherOrder}. Interestingly, this expansion shows that the inclusion of quantum coherence cannot generically change the ballistic nature of transport in finite-range, noninteracting systems. In this paper, we therefore focus on the ballistic or Euler scaling limit~\cite{spohn2012large} as $\|\bx\|,t \to \infty$. This is precisely the regime of validity of the usual free-streaming Boltzmann equation. In this limit, terms of order $O(\partial_{\bx}^2)$ in spatial derivatives can be neglected, and taking expectation values of the operator equations of motion with respect to the initial density matrix yields the transport equation
\begin{align}
    \partial_t \,w(\bx,\bk,t) = &-i\big[h(\bk),w(\bx,\bk,t)\big]\nonumber\\ 
    &
    -\frac{1}{2}\sum_{a=1}^d\big\{\partial_{k_a} h(\bk),\partial_{x_a} w(\bx,\bk,t) \big\}\,.
    \label{eqn:EoM}
\end{align}
for the matrix-valued orbital Wigner function $w$. 

Substituting the approximate Wigner function $w(\bx,\bk,t)$ resulting from solving Eq.~\eqref{eqn:EoM} into Eqs.~\eqref{eq:Wignerpredn} and \eqref{eq:Wignerpredj} yields ballistic-scale transport-theory predictions for $n(\bR_j,t)$ and $\textbf{j}(\bR_j,t)$, which we denote by $n_{\mathrm{Wigner}}(\bx,t)$ and $\textbf{j}_{\mathrm{Wigner}}(\bx,t)$ respectively. For the extension of the equation of motion to static electric fields we refer to Appendix~\ref{appendix:extfield}.

We note that Eq.~\eqref{eqn:EoM} appeared previously in the literature as Eq.~(9) of the supplementary data of Ref.~\cite{Cepellotti2021}, though the final result of that paper is inequivalent to ours. An equivalent kinetic equation to ours for (bosonic) phonons was proposed in Ref.~\cite{Simoncelli2022}. While we focus exclusively on fermions in this paper, we note that the derivation of Eq.~\eqref{eqn:EoM} for bosons is formally identical to the derivation presented here and differs from fermions only in the specification of a locally thermal initial state, Eq.~\eqref{eq:locthermwigner}, which must be replaced by a local Bose-Einstein distribution.

\subsection{Initial conditions}
\label{sec:IC}
For any initial density matrix $\hat{\rho}$, the exact initial condition for the Wigner function $w_{\alpha \beta}$ is given in principle by Eq.~\eqref{eqn:OrbitalWignerFunction}. However, this expression can be cumbersome to evaluate in practice and it will be useful to instead introduce approximate but easy-to-calculate initial conditions for $w_{\alpha \beta}$ that model the local equilibrium states of interest in transport theory. Recall that slowly varying initial temperature and chemical potential profiles $\beta(\bx)$ and $\mu(\bx)$ define local equilibrium states of the form
\begin{equation}
    \label{eq:loctherm}
    \hat{\rho} \propto \exp{\Bigg(-\sum_{j\in\Lambda} \beta(\bR_j)\Big[\hat{h}_j - \mu(\bR_j)\, \hat{n}_j\Big]\Bigg)}\,,
\end{equation}
where $\hat{h}_j$ and $\hat{n}_j$ denote the energy and charge density operators defined in Eqs.~\eqref{eqn:localHamiltonian} and \eqref{eqn:localCharge} respectively. To model these, let us first consider the Wigner function for initial states in global thermal equilibrium $\hat{\rho} \propto e^{-\beta (\hat{H}_0-\mu \hat{N})}$. In terms of the changes of basis to Bloch electrons $U(\mathbf{k})$ defined in Eq.~\eqref{eq:changeofbasis}, the exact result
\begin{align}
    w_{\alpha \beta}(\bk) = \sum_{n=1}^M U_{\alpha n}(\bk) \frac{1}{1+e^{\beta(E_n(\bk)-\mu)}} U^\dagger_{n \beta}(\bk)
\end{align}
follows by Fermi-Dirac statistics. By definition of $U(\bk)$, this can be written in matrix form in terms of the Bloch Hamiltonian $h(\bk)$ as
\begin{equation}
    w(\bk) = \frac{1}{1+e^{\beta(h(\bk)-\mu)}}\,.
\end{equation}
In the local density approximation (whose rigorous justification requires substantial effort~\cite{fagotti2024asymptoticbehaviourdeterminantsexpansion} and will not be pursued here) and in the same notation, the initial condition corresponding to Eq.~\eqref{eq:loctherm} can be written as
\begin{align}
    w(\bx,\bk,0) = \frac{1}{1+e^{\beta(\bx)[h(\bk)-\mu(\bx)]}}\,,
    \label{eq:locthermwigner}
\end{align}
which is expected to be accurate whenever the length scale of variation of $\beta(\bx)$ and $\mu(\bx)$ is much larger than the typical correlation length of the state in Eq.~\eqref{eq:loctherm}. More precisely, we expect that provided $\max_{\bx} \beta(\bx) < \infty$, Eq.~\eqref{eq:locthermwigner} can be made to hold to an arbitrarily good approximation by taking the scale of variation of $\beta(\bx)$ and $\mu(\bx)$ to be sufficiently large. We will use this expression for comparison with microscopic numerical simulations of time evolution from Eq.~\eqref{eq:loctherm} below.

We will also be interested in more strongly nonequilibrium initial conditions that consist of preparing a local equilibrium state in Eq.~\eqref{eq:loctherm} and then applying an inhomogeneous and slowly varying phase texture $\varphi_\alpha(\bx)$ in space  $\hat{c}_{j\alpha} \mapsto e^{-i\varphi_\alpha(\bR_{j\alpha})}\hat{c}_{j\alpha}$. Such initial states can in principle be prepared and measured using present-day experimental capabilities~\cite{Santra_2017,murthy2019directimagingorderparameter,bruggenjurgen2024phase,yang2019coherent}. Let us write
\begin{equation}
    \label{eq:rotlocham}
    \hat{h}'_j = -\sum_{j'\in \Lambda}\sum_{\alpha,\beta=1}^M t^{\alpha\beta}_{jj'}\,\, e^{i\big(\varphi_{\alpha}(\bR_{j\alpha})-\varphi_{\beta}(\bR_{j'\beta})\big)}\,\hat{c}^\dagger_{j\alpha}\,\hat{c}^{}_{j'\beta}
\end{equation}
for the modified local Hamiltonian under this mapping. Then the initial condition in question is given explicitly by
\begin{equation}
    \label{eq:phasecoh}
    \hat{\rho}' \propto \exp{\Bigg(-\sum_{j\in\Lambda} \beta(\bR_j)\big[\hat{h}'_j - \mu(\bR_j) \,\hat{n}_j\big]\Bigg)}\,.
\end{equation}
In the local density approximation, the initial Wigner function in Eq.~\eqref{eq:locthermwigner} modeling Eq.~\eqref{eq:phasecoh} is modified to 
\begin{equation}
    w_{\alpha \beta}(\mathbf{x},\mathbf{k},0) \mapsto e^{i[\varphi_\alpha(\bx) - \varphi_\beta(\bx)]}\,w_{\alpha \beta}(\bx,\bk,0) 
\end{equation}
at leading order in spatial derivatives $\partial_\bx \varphi_{\alpha}$. A nontrivial probe of such initial phase coherence is the two-point correlation function $\hat{c}_{j\alpha}^\dagger \hat{c}^{}_{j\beta}$ between two basis sites within the same unit cell. We can extract this information from the Wigner function using the identity
\begin{align}
    \int_{\mathrm{BZ}} \frac{d\mathbf{k}}{V_{\mathrm{BZ}}} e^{-i\bk \cdot (\btau_\alpha - \btau_{\beta})} \,w_{\beta \alpha}[&\bR_j \!+\! (\btau_\alpha\!+\!\btau_\beta)/2,\bk]
    \label{eq:phasecohmarg}
    = \langle \hat{c}_{j\alpha}^\dagger \hat{c}^{}_{j\beta}\rangle\,.
\end{align}

\subsection{Time-dependent Hamiltonians}
We now consider allowing the tight-binding Hamiltonian Eq.~\eqref{eqn:Hamiltonian} to depend on time via its hopping amplitudes $t_{\alpha \beta}(\br,t)$. Such an extension of our formalism is necessary to capture topological effects such as Thouless pumping~\cite{thouless1983quantization}. In Fourier space, this yields the time-dependent Hamiltonian
\begin{equation}
    \hat{H}_0(t) = \int_{\mathrm{BZ} }\frac{d\mathbf{k}}{V_{\mathrm{BZ}}} \sum_{\alpha,\beta=1}^M h_{\alpha \beta}(\mathbf{k},t) \,\hat{c}_{\alpha}^\dagger(\bk)\,\hat{c}^{}_{\beta}(\bk)\,,
\end{equation}
where the time-dependent Bloch Hamiltonian is
\begin{equation}
    \label{eq:BlochHamiltonian}
    h_{\alpha \beta}(\mathbf{k},t) = -\sum_{j\in\Lambda} t_{\alpha \beta}(\bR_j,t)\, e^{-i\mathbf{k}\cdot (\bR_j +\btau_{\alpha}-\btau_\beta)}\,.
\end{equation}
The change of basis to Bloch electrons now depends on time, i.e. for each $\mathbf{k}$, there exists $U(\mathbf{k},t)$ such that 
\begin{equation}
    \sum_{\alpha,\beta=1}^M U^{\dagger}_{m \alpha}(\mathbf{k},t) \,h_{\alpha \beta} (\mathbf{k},t)\,U_{\beta n}(\mathbf{k},t) = E_{n}(\mathbf{k},t)\, \delta_{mn}\,,
\end{equation}
which defines $M$ time-dependent energy bands $\{E_n(\mathbf{k},t)\}_{n=1}^M$ and time-dependent Bloch electron creation operators
\begin{equation}
    \hat{d}^\dagger_{n}(\mathbf{k},t) = \sum_{\alpha=1}^M U_{\alpha n}(\mathbf{k},t)\,\hat{c}^\dagger_{\alpha}(\bk)\,,
\end{equation}
in terms of which
\begin{equation}
    \hat{H}_0(t) = \int_{\mathrm{BZ}} \frac{d\mathbf{k}}{V_{\mathrm{BZ}}} \sum_{n=1}^M E_n(\mathbf{k},t) \,\hat{d}^\dagger_n(\mathbf{k},t) \,\hat{d}^{}_n(\mathbf{k},t)\,.
\end{equation}
A special case that admits further simplification is periodic-in-time or Floquet dynamics $\hat{H}_0(t+T) = \hat{H}_0(t)$, see Refs.~\cite{Genske_2015, Esin} for proposed Floquet-Boltzmann equations that capture such dynamics. (However, the authors of these works focus on diagonal elements of the Wigner function in the basis of Bloch electrons, which is distinct from the approach pursued here.) Such simplifications arising from periodicity in time will not be necessary for our purposes and our transport equation instead allows for arbitrary time dependence, which both encompasses the time-independent case discussed above and matches microscopic numerical simulations in the time-periodic case (see Sec.~\ref{sec:timedependence}). 

By repeating the derivation of Eq.~\eqref{eqn:EoM} in the time-dependent setting, we obtain the ballistic-scale transport equation
\begin{align}
    \partial_t w(\bx,\bk,t) = &-i\big[h(\bk,t),w(\bx,\bk,t)\big]\nonumber\\ 
    &-\frac{1}{2}\sum_{a=1}^d\big\{\partial_{k_a} h(\bk,t),\partial_{x_a} w(\bx,\bk,t) \big\}\,.
    \label{eqn:tdepWignerEoM}
\end{align}
We note that this time-dependent equation can also be used to capture the effect of spatially uniform electrostatic fields upon changing to a uniformly accelerating frame of reference; see Appendix~\ref{appendix:extfield}.

\subsection{Methods of solution}
\subsubsection{General considerations}
The transport Eq.~\eqref{eqn:EoM} and its time-dependent generalization Eq.~\eqref{eqn:tdepWignerEoM} are linear, first-order PDEs and therefore always amenable to a numerical solution based on finite-difference methods~\cite{press2007numerical} or Fourier analysis. We pursue both approaches in this work and two possible numerical schemes are summarized in Appendix \ref{appendix:numericalmethods}.

A striking and immediate property of these equations, shared by other Wigner-function based equations of motion~\cite{demeio2002wigner,unlu2004multi,Culcer2006,Culcer2009,Morandi2009,Wong2011,Wickles2013,Morawetz2015,Iafrate2017,Sekine2017,Cepellotti2021,Bhalla2021,Konig2021,Valet2023}, is that they exhibit $M^2$ propagation velocities at any given instant of time, rather than the $M$ quasiparticle group velocities $\{\partial_{\bk}E_n(\bk)\}_{n=1}^M$ within each band that determine ordinary semiclassical transport. We will explore this observation in more detail below.

For now, we merely note that the smaller the number of bands $M$, the more tractable Eq.~\eqref{eqn:EoM} becomes. The simplest cases are $M=1$ (a single band), for which the orbital Wigner function recovers conventional Boltzmann theory at leading order in spatial derivatives~\cite{fagotti2017higher,essler2023short}, and $M=2$ bands, which is the simplest case that exhibits effects beyond Boltzmann theory and will be the main focus of this paper.

For the two-band systems considered in this paper, we obtain predictions for orbital Wigner function dynamics by solving Eq.~\eqref{eqn:EoM} numerically on a sufficiently dense grid of $k$ points in the BZ, using a second-order staggered leapfrog method~\cite{press2007numerical} for Figs. \ref{Fig:BoltzmannComp}--\ref{fig:ThoulessTest} and a more efficient GPU-based Fourier-kernel method for the most numerically demanding Fig. \ref{fig:FlatBands}, see Appendix \ref{appendix:numericalmethods} for further details.

\subsubsection{Solvable special case: Vanishing Berry connection}
In general, we expect that the nontrivial $\bk$ dependence of the Bloch eigenvectors [i.e. of the matrix $U(\bk)$ defined in Eq.~\eqref{eq:changeofbasis}] precludes a simple exact solution to Eq.~\eqref{eqn:EoM}. In the special case that $U(\bk)$ is differentiable in $\bk$ and defines a vanishing Berry connection $\mathcal{A}_a(\bk) = -i U(\bk)^\dagger \partial_{k_a} U(\bk) = 0$, such a solution is possible. Since its derivation is instructive and because it may be found useful for benchmarking numerical approaches, we now summarize this closed-form solution.

First define the {\it band Wigner function} 
\begin{align}
    \label{eqn:bandWigner}
    \tilde{w}(\bx,\bk,t) = U^\dagger(\bk)\, w (\bx,\bk,t)\,U(\bk),
\end{align}
(this is analogous to the decoupling step in Refs.~\cite{Wong2011, Wickles2013}) and the diagonalized Bloch Hamiltonian $\tilde{h}(\bk) = U^\dagger(\bk)\, h (\bk)\,U(\bk)$ as in Eq.~\eqref{eq:changeofbasis}. Then substituting into Eq.~\eqref{eqn:EoM} and suppressing arguments, we find that
\begin{equation}
    \label{eq:unsimpbandeq}
    \partial_t \tilde{w} + \frac{1}{2}\sum_{a=1}^d \big\{U^\dagger\,(\partial_{k_a} h)\, U,\partial_{x_a}\tilde{w}\big\} = -i\big[\tilde{h}, \tilde{w}\big]\,.
\end{equation}
To simplify the streaming term, note that by differentiability of $U(\bk)$, $\partial_{k_a} U(\bk) = - U^\dagger(\bk) \,[\partial_{k_a}U(\bk)]\, U^\dagger(\bk)$ and therefore
\begin{equation}
    \label{eq:BerryConnAppears}
    U^\dagger(\bk) \,\big[\partial_{k_a} h(\bk)\big]\, U(\bk) = \partial_{k_a} \tilde{h}(\bk) + i\big[\mathcal{A}_a(\bk),\tilde{h}(\bk)\big]\,,
\end{equation}
where we defined the Berry connection through its components $\mathcal{A}_a(\bk) = -i U^\dagger(\bk)\, \partial_{k_a} U(\bk)$. Restoring matrix indices, it follows that
\begin{align}
    \nonumber
    &\partial_t \tilde{w}_{mn} + \frac{1}{2}\sum_{a=1}^d \Big(\big\{\partial_{k_a} \tilde{h} +i\big[\mathcal{A}_a,\tilde{h}\big],\partial_{x_a}\tilde{w}\big\}\Big)_{mn} \\
    \label{eq:simpbandeq}
    &= -i(E_m-E_n)\tilde{w}_{mn}\,.
\end{align}
Although this equation is always exactly solvable by linearity, the resulting solution does not appear to be especially tractable in general. An exception is the special case $\mathcal{A}_a(\bk)=0$ for all $\bk$, in which Eq.~\eqref{eq:simpbandeq} reduces to
\begin{equation}
    \label{eq:idealbandeq}
    \partial_t \tilde{w}_{mn} + \bar{\mathbf{v}}_{mn} \cdot \partial_{\bx}\tilde{w}_{mn} = -iE_{mn}\tilde{w}_{mn}\,,
\end{equation}
where we defined $E_{mn}(\bk) := E_m(\bk)-E_n(\bk)$ and a mean group velocity $\bar{\mathbf{v}}_{mn}(\bk) = \frac{1}{2}\partial_{\bk}[E_m(\bk)+E_n(\bk)]$. Equation~\eqref{eq:idealbandeq} is now simple enough to be solved in closed form, and the solution to its initial value problem reads
\begin{equation}
    \tilde{w}_{mn}(\bx,\bk,t) = e^{-iE_{mn}(\bk)t} \tilde{w}_{mn}[\bx-\bar{\mathbf{v}}_{mn}(\bk) t,\bk,0]\,.
\end{equation}
In terms of the orbital Wigner function, the resulting solution to Eq.~\eqref{eqn:EoM} is given by
\begin{align}
    \nonumber
    w_{\alpha \beta}(\bx,&\bk,t) = \sum_{\alpha',\beta',m,n=1}^M e^{-iE_{mn}(\bk)t} U_{\alpha m}(\bk) U^\dagger_{m\alpha'} (\bk) \\
    &\times w_{\alpha'\beta'}(\bx-\bar{\mathbf{v}}_{mn}(\bk)t,\bk,0) U_{\beta'n}(\bk) U^\dagger_{n\beta}(\bk)\,.
\end{align}

\section{Comparison with semiclassical transport theory}
\label{sec:ComparsionSemiClassic}

We now test the validity of the orbital Wigner function approach compared with conventional semiclassical transport theory, against numerically exact free-fermion simulations. We first consider states that are locally in thermal equilibrium at a nonzero temperature, for which we do not expect any deviations from textbook results~\cite{ashcroft1976solid}. We present a general analytical argument for how orbital Wigner functions recover semiclassical transport theory in local thermal equilibrium before illustrating this point through a free-fermion simulation of charge and current dynamics in a two-band example.

We next consider initial states with spatially inhomogeneous single-particle quantum coherence, which lie beyond the scope of semiclassical transport theory. We specifically focus on quantum coherence between orbitals within each unit cell, as might be prepared in an ultracold atom device~\cite{Santra_2017,murthy2019directimagingorderparameter,bruggenjurgen2024phase} or through scanning tunneling microscopy of a material sample~\cite{yang2019coherent}. We present numerical evidence that orbital Wigner functions are well suited for capturing relaxation from such quantum-coherent initial states.

\subsection{Essentials of semiclassical transport theory}

For generic choices of hoppings, we may assume that the energy bands do not intersect $E_1(\mathbf{k}) > E_2(\mathbf{k}) \ldots > E_M(\mathbf{k})$ (note that this assumption is not needed for our orbital Wigner function formalism but rather to facilitate comparison with conventional transport theory). Then defining group velocities $\mathbf{v}_{n}(\mathbf{k}) = \partial_{\mathbf{k}} E_n(\mathbf{k})$, 
conventional semiclassical transport theory in the absence of interactions and external fields describes the time evolution of the local density of occupied states $\rho_n(\mathbf{x},\mathbf{k},t)$ of the $n$th band via the Boltzmann equation~\cite{ashcroft1976solid} 
\begin{equation}
    \label{eq:Boltzeq}
    \partial_t \rho_n + \mathbf{v}_n(\mathbf{k}) \cdot \partial_{\mathbf{x}} \rho_n = 0\,,
\end{equation}
with solution
\begin{equation}
    \rho_n(\mathbf{x},\mathbf{k},t) = \rho_n[\mathbf{x}-\mathbf{v}_n(\bk)t,\mathbf{k},0]\,.
\end{equation}
In thermal equilibrium states, $\hat{\rho} \propto e^{-\beta (\hat{H}_0-\mu \hat{N})}$, we have
\begin{equation}
    \rho_n(\bk) =  \frac{1}{1+e^{\beta[E_n(\bk)-\mu]}}\,,
\end{equation}
which in the local density approximation yields the initial condition
\begin{equation}
    \label{eq:locthermBoltz}
    \rho_n(\bx,\bk,0) = \frac{1}{1+e^{\beta(\bx)[E_n(\bk)-\mu(\bx)]}}
\end{equation}
for locally thermal initial states in Eq.~\eqref{eq:loctherm}. Predictions for the dynamics of local observables such as the charge density $n(\bx,t)$ and charge-current density $\mathbf{j}(\bx,t)$ are then given by 
\begin{equation}
    \label{eq:boltzmannpredn}
    n_{\mathrm{Boltzmann}}(\bx,t) = \int_{\mathrm{BZ}} \frac{d\bk}{V_{\mathrm{BZ}}} \, \sum_{n=1}^M \rho_n(\bx,\bk,t) \, ,
\end{equation} 
and
\begin{equation}
    \label{eq:boltzmannpredj}
    \mathbf{j}_{\mathrm{Boltzmann}}(\bx,t) = \int_{\mathrm{BZ}} \frac{d\bk}{V_{\mathrm{BZ}}} \,  \sum_{n=1}^M \rho_n(\bx,\bk,t) \mathbf{v}_n(\bk) \, ,
\end{equation}
respectively.

\subsection{Dynamics of local equilibrium states}
\label{sec:loceq}

\begin{figure}[t]
    \centering 
\includegraphics[width=0.99\linewidth]{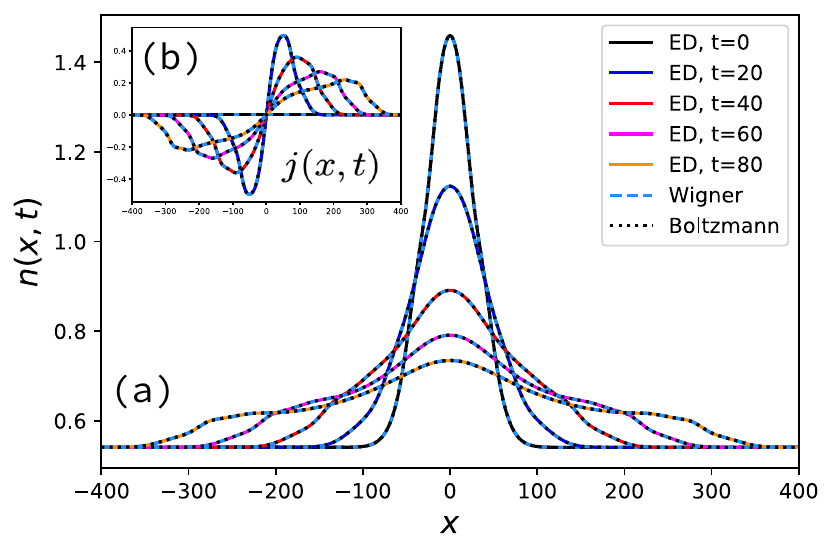}
    \caption{(a) Time evolution from a localized initial charge density under the Hamiltonian in Eq.~\eqref{eq:hamchoice}, with uniform initial temperature $\beta=1$ and initial chemical potential profile as in Eq.~\eqref{eq:chargespot}, with bulk chemical potential $\mu_0=-4$, central chemical potential $\mu_1=4$ and length scale $l=40$ sites. The agreement between microscopic exact diagonalization (solid lines), Wigner function predictions (light blue dashed lines) and Boltzmann equation predictions (black dotted lines) is clear. (b) Predictions for the charge-current density $j(x,t)$ vs numerics.}
    \label{Fig:BoltzmannComp}
\end{figure}

We commented earlier that the equations of motion for the orbital Wigner function given in Eq.~\eqref{eqn:EoM} were characterized by $M^2$ propagation velocities rather than the $M$ quasiparticle group velocities that appear in the semiclassical theory. On the other hand, we expect the semiclassical theory to yield accurate results near thermal equilibrium, for example for time evolution from local equilibrium initial states of the form in Eq.~\eqref{eq:loctherm}.

In fact there is no contradiction between these claims, and in hydrodynamic regimes for which the system is locally near thermal equilibrium, the orbital Wigner dynamics closely approximates free Boltzmann dynamics and is dominated by $M$ diagonal degrees of freedom that propagate at the quasiparticle group velocity.
In this section, we first verify analytically that this claim holds instantaneously for locally equilibrated Wigner functions of the form in Eq.~\eqref{eq:locthermwigner}. We then check numerically that the time evolution of charge density from local equilibrium initial conditions in Eq.~\eqref{eq:loctherm} is accurately captured by the Wigner-function transport Eq.~\eqref{eqn:EoM} in a two-band example.

For the analytical argument, we take as our starting point the transport Eq.~\eqref{eq:simpbandeq} for the band Wigner function. For an orbital Wigner function that is instantaneously in local equilibrium, so that
\begin{equation}
    \label{eq:loceqansatz}
    w(\bx,\bk,t) = \frac{1}{1+e^{\beta(\bx,t)[h(\bk)-\mu(\bx,t)]}}\,,
\end{equation}
the band Wigner function is diagonal
\begin{equation}
    \tilde{w}_{mn}(\bx,\bk,t) = \frac{\delta_{mn}}{1+e^{\beta(\bx,t)[E_n(\bk)-\mu(\bx,t)]}}\,.
\end{equation}
Thus the off-diagonal components of Eq.~\eqref{eq:simpbandeq} can be ignored in the vicinity of time $t$, leaving the evolution equation
\begin{align}
    \nonumber
    \partial_t \tilde{w}_{nn} + \frac{1}{2}\sum_{a=1}^d \Big(\big\{\partial_{k_a} \tilde{h} +i\big[\mathcal{A}_a,\tilde{h}\big],\partial_{x_a}\tilde{w}\big\}\Big)_{\!nn} = 0
\end{align}
for the diagonal components. Evaluating the anticommutator directly and using the fact that $\tilde{h}$ is diagonal then yields
\begin{equation}
    \partial_t \tilde{w}_{nn}(\bx,\bk,t) + \mathbf{v}_n(\bk) \cdot \partial_\mathbf{x} \tilde{w}_{nn}(\bx,\bk,t) = 0\,.
\end{equation}
Note that this perfectly recovers the standard Boltzmann description of time evolution from local equilibrium, provided we make the identification $\tilde{w}_{nn}(\mathbf{x},\mathbf{k},t) \leftrightarrow \rho_n(\mathbf{x},\mathbf{k},t)$.

In practice, the local equilibrium ansatz in Eq.~\eqref{eq:loceqansatz} is not preserved exactly under the orbital Wigner function dynamics, and for a more stringent test, we now compare predictions from solving Eq.~\eqref{eqn:EoM} with the initial condition in Eq.~\eqref{eq:locthermwigner} to microscopic numerical simulations from locally thermal initial conditions in Eq.~\eqref{eq:loctherm} in a specific two-band model. 

For our numerical comparison, we choose the 1D two-band Hamiltonian
\begin{align}
    \label{eq:hamchoice}
    \hat H_0 = \sum_{j=-L/2}^{L/2-1}\,\big(4\,\hat c^\dagger_{j,A}\hat c^{}_{j,B}-3\,\hat c^\dagger_{j+1,A}\hat c^{}_{j,A}+\text{H.c.}\big)\,.
\end{align}
This has Bloch Hamiltonian 
\begin{align}
    h(k) = \begin{pmatrix} -6 \cos k & 4 \\ 4 & 0 \end{pmatrix}\,.
\end{align}
The band structure is given by $E_{\pm}(k) = - 3 \cos{k} \pm \sqrt{16 + 9 \cos^2{k}}$ and the Bloch eigenvectors are supported on both sites $A$ and $B$ and depend nontrivially on $k$. For mathematical convenience, we set the basis positions $\btau_{A} = \btau_B = 0$ within each unit cell and index lattice sites by integers, $x_j = j$ with $-L/2 \leq x_j \leq L/2$ and periodic boundary conditions $x_{-L/2} \equiv x_{L/2}$. 

For this model, the Heisenberg-picture charge density operator $\hat{n}_j = \hat{c}_{j,A}^\dagger\hat{c}^{}_{j,A}+\hat{c}_{j,B}^\dagger\hat{c}^{}_{j,B}$ satisfies the microscopic continuity equation
\begin{equation}
    \label{eq:continuity}
    \partial_t \hat{n}_j + \hat{j}_{j+1}-\hat{j}_j = 0\,,
\end{equation}
where the charge-current density operator $\hat{j}_j = 3i(\hat{c}_{j,A}^\dagger \hat{c}^{}_{j-1,A}-\hat{c}_{j-1,A}^\dagger \hat{c}^{}_{j,A})$ yields a global current operator $\hat{J}=\sum_{j=-L/2}^{L/2-1} \hat{j}_j$ that satisfies Eq.~\eqref{eq:defJ} in the thermodynamic limit. 

We consider initializing this model in a local equilibrium state as in Eq.~\eqref{eq:loctherm} at a uniform temperature $\beta=1$ but with a strongly inhomogeneous chemical potential that creates a localized distribution of charge 
\begin{equation}
    \label{eq:chargespot}
    \mu(x) = \mu_0 + (\mu_1-\mu_0)e^{-x^2/l^2}
\end{equation}
about $x=0$, where the bulk chemical potential $\mu_0 = - 4$ sits in the lower band, the central chemical potential $\mu_1 = 4$ sits in the upper band, and the length scale $l=40$. Note that although this state is locally in thermal equilibrium, it is far from the linear response regime and thus represents a stringent test of the theory.

We perform numerically exact microscopic simulations using standard free-fermion methods~\cite{cheong2004many,peschel2003calculation} on $L=800$ sites and compare the resulting dynamics of the charge density and charge-current density to Boltzmann equation predictions in Eqs.~\eqref{eq:boltzmannpredn} and \eqref{eq:boltzmannpredj} and to the Wigner-function predictions in Eq.~\eqref{eq:Wignerpredn} and \eqref{eq:Wignerpredj} obtained by solving Eq.~\eqref{eqn:EoM} numerically. 
Results are shown in Fig.~\ref{Fig:BoltzmannComp} and the charge and current profiles arising from these three distinct methods are indistinguishable from one another as expected.

\begin{figure*}[t]
    \centering
\includegraphics[width=0.99\linewidth]{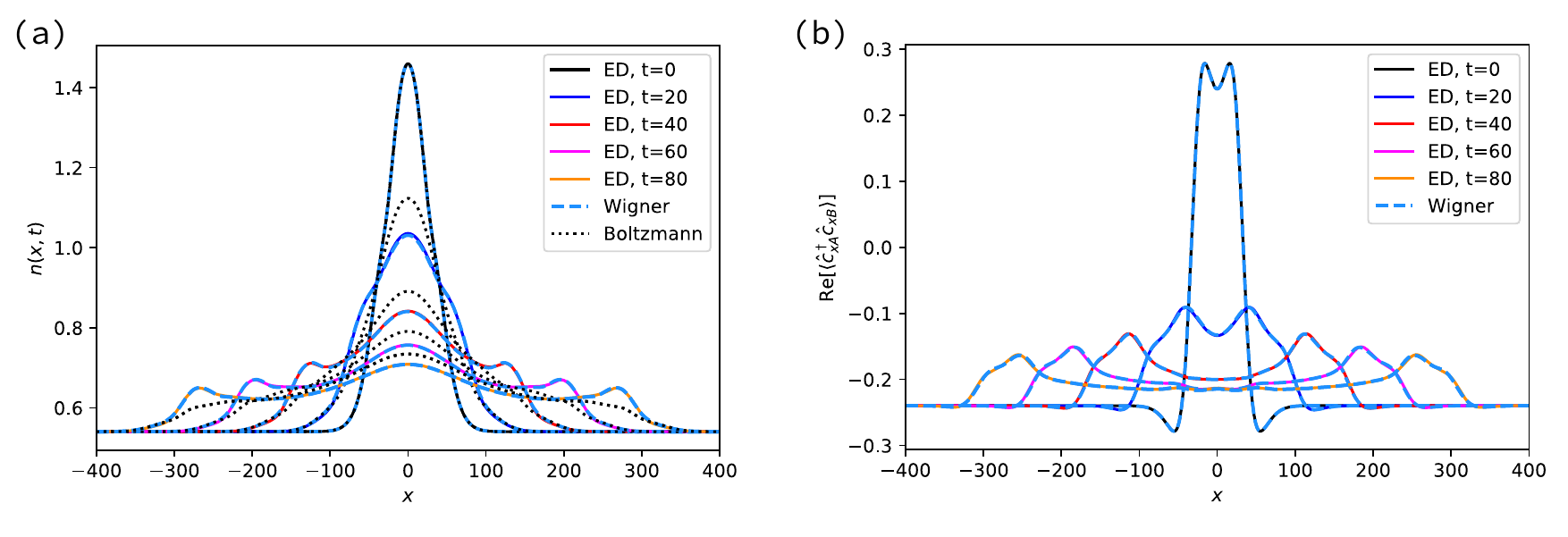}
    \caption{(a) Time evolution of charge density from the initial condition in Eq.~\eqref{eq:chargespot} with the same parameters as in Fig.~\ref{Fig:BoltzmannComp}, augmented by an inhomogeneous initial phase as in Eq.~\eqref{eq:inhomphase} with maximum relative phase $\phi=\pi$. Boltzmann theory predictions that do not account for this inhomogeneous phase (as plotted in Fig.~\ref{Fig:BoltzmannComp}) appear as black dotted lines for reference. (b) Wigner function predictions for the dynamics of onsite inter-orbital correlation functions. Such correlation functions are beyond conventional semiclassical transport theory and we observe excellent agreement with Wigner function predictions.}
    \label{Fig:PhaseCohCompn}
\end{figure*}

\subsection{Phase coherence beyond semiclassical transport}

We next test the orbital Wigner function approach for phase-coherent initial conditions of the form in Eq.~\eqref{eq:phasecoh}. Textbook semiclassical transport theory does not make any predictions for such dynamics~\cite{ashcroft1976solid} because the initial state in Eq.~\eqref{eq:phasecoh} is not even locally in thermal equilibrium with respect to $\hat{H}_0$. Note however that this state is in local equilibrium with respect to the effective Hamiltonian $\hat{H}'_0 = \sum_{j \in \Lambda} h'_j$ with $h'_j$ defined as in Eq.~\eqref{eq:rotlocham}. (In this sense, Eq.~\eqref{eq:phasecoh} defines a kind of quantum quench~\cite{mitra2018quantum}.)

In this setting, our transport Eq.~\eqref{eqn:EoM} makes predictions both for the dynamics of the local charge density $n(x,t)$ and for the dynamics of local off-diagonal correlation functions $\langle \hat{c}_{j\alpha}^\dagger \hat{c}^{}_{j\beta}\rangle(t) = \mathrm{Tr}[\hat{\rho}(t) \hat{c}_{j\alpha}^\dagger\hat{c}^{}_{j\beta}]$ via Eq.~\eqref{eq:phasecohmarg}, which here yields the prediction
\begin{equation}
    \langle \hat{c}_{j\alpha}^\dagger \hat{c}^{}_{j\beta}\rangle_{\mathrm{Wigner}}(t) = 
    \int_{\mathrm{BZ}} \frac{dk}{V_{\mathrm{BZ}}} w_{\beta\alpha}(x_j,k,t)\,.
\end{equation}
We test this prediction for the Hamiltonian in Eq.~\eqref{eq:hamchoice} considered in the previous section, with the same initial conditions consisting of a localized spot of charge as in Eq.~\eqref{eq:chargespot} but now augment these by a relative phase with maximal value $\phi$ that varies spatially on the same length scale as the charge density, i.e.,
\begin{equation}
    \label{eq:inhomphase}
    \varphi_A(x) = (\phi/2) e^{-x^2/l^2}, \quad \varphi_B(x) = -(\phi/2) e^{-x^2/l^2}\,,
\end{equation}
giving rise to a strongly non-equilibrium initial density matrix $\hat{\rho}'$ as defined in Eq.~\eqref{eq:phasecoh}.

For such initial conditions, we find excellent agreement with orbital Wigner function predictions for both $n(x,t)$ and for the onsite correlation functions $\langle \hat{c}_{j\alpha}^\dagger \hat{c}^{}_{j\beta} \rangle(t)$, see Fig.~\ref{Fig:PhaseCohCompn} for a comparison with microscopic simulations with $\phi=\pi$. Note that the inhomogeneous initial phase in Eq.~\eqref{eq:inhomphase} does not alter the initial charge-density profile but it does alter the initial energy density with respect to the Hamiltonian $\hat{H}_0$. This is one way to understand the deviation in the left panel of Fig.~\ref{Fig:PhaseCohCompn} from Boltzmann equation predictions for the charge dynamics that do not consider the inhomogeneous initial phase in Eq.~\eqref{eq:inhomphase}.

\section{Application to time-dependent Hamiltonians}
\label{sec:timedependence}

Coherent driving of quantum Hamiltonians is well known to induce coherent superpositions of quantum states; the canonical example is Rabi oscillations in two-level systems. It is therefore natural to expect that quantum coherence plays an essential role in the dynamics and transport of time-dependent multiband systems. Indeed, the adiabatic Thouless pumping of a topologically quantized unit of charge per cycle in 1D insulators can be seen as a consequence of the states within a topological band accumulating a coherent global phase~\cite{thouless1983quantization}.

In this section, we perform a precision verification of the accuracy of our orbital Wigner formalism for time-dependent Hamiltonians against microscopic free-fermion simulations, which to the best of our knowledge has not been performed for existing generalizations of the Boltzmann equation to time-dependent settings~\cite{Genske_2015,Esin}. We note that our approach requires no assumptions on the time-periodicity of the Hamiltonian or the adiabaticity of the drive.

\subsection{Pumping local equilibrium states}

\begin{figure}[t]
    \centering
\includegraphics[width=0.99\linewidth]{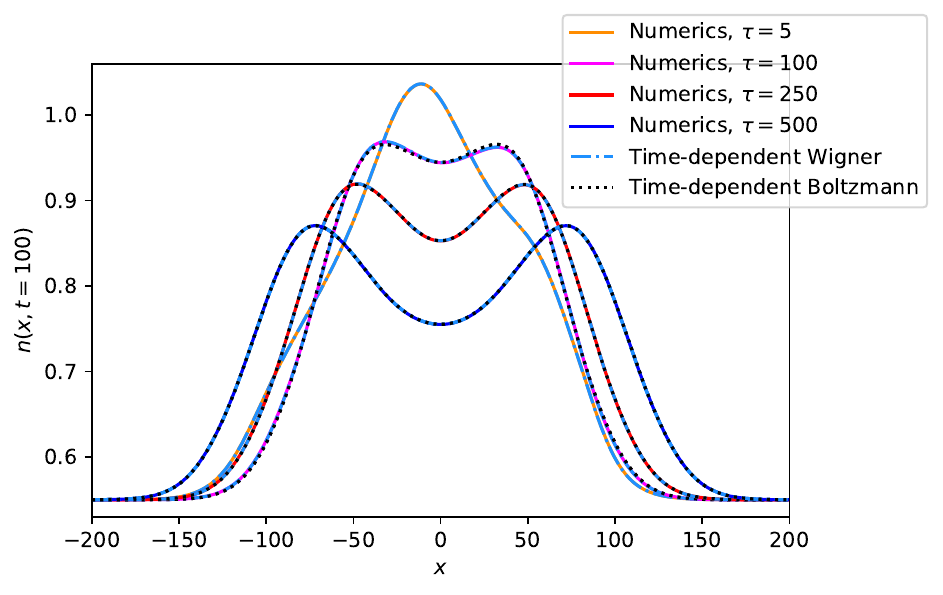}
\caption{Numerical time evolution of an initially localized charge density under the periodic time evolution in Eq.~\eqref{eq:tdephamchoice} after a time $t=100$, for different driving periods $\tau = 5,100,250,500$, compared with time-dependent Wigner function predictions (light blue dashed lines) and time-dependent Boltzmann predictions (black dotted lines), see text for further details. The time-dependent Boltzmann equation works well for timescales $\tau \gtrsim t$ but freezes for $\tau \lesssim t$, ceasing to provide an accurate description of the dynamics (the Boltzmann predictions for $\tau=100$ and $5$ are indistinguishable by eye). In contrast, the time-dependent Wigner function yields extremely accurate results over two orders of magnitude of driving periods.}
    \label{fig:loceqtdep}
\end{figure}

The na{\"i}ve generalization of the Boltzmann Eq.~\eqref{eq:Boltzeq} to the time-dependent case reads
\begin{equation}
    \label{eq:tdepBoltz}
    \partial_t \rho_n + \mathbf{v}_n(\mathbf{k},t) \cdot \partial_{\mathbf{x}} \rho_n = 0\,,
\end{equation}
where the group velocities for each band $\mathbf{v}_n(\mathbf{k},t) := \partial_{\mathbf{k}} E_n(\mathbf{k},t)$ now depend on time; initial data for local equilibrium states is again specified by Eq.~\eqref{eq:locthermBoltz} and predictions for the dynamics of the local charge density $n(\mathbf{x},t)$ are again obtained from Eq.~\eqref{eq:boltzmannpredn}. Analogous equations with $\mathbf{v}_n$ replaced by a time-dependent quasiparticle velocity have been found to perform well for time-dependent and instantaneously quantum integrable Hamiltonians in one spatial dimension~\cite{Bastianello_2019}.

We can justify this prescription microscopically and in local equilibrium by repeating the arguments of Sec.~\ref{sec:loceq}. We find that starting from a local equilibrium state at time $t$, Eq.~\eqref{eqn:tdepWignerEoM} implies that the diagonal components of the band Wigner function $\tilde{w}(\bx,\bk,t) = U^\dagger(\bk,t) \,w(\bx,\bk,t) \,U(\bk,t)$ evolve instantaneously according to the time-dependent Boltzmann equation
\begin{equation}
    \partial_t \tilde{w}_{nn}(\bx,\bk,t) + \mathbf{v}_n(\bk,t) \cdot \partial_\mathbf{x} \tilde{w}_{nn}(\bx,\bk,t) = 0\,.
\end{equation}
As in the time-independent case, the na{\"i}ve Boltzmann prediction Eq.~\eqref{eq:tdepBoltz} is recovered once we make the identification $\tilde{w}_{nn}(\mathbf{x},\mathbf{k},t) \leftrightarrow \rho_n(\mathbf{x},\mathbf{k},t)$.

By the method of characteristics~\cite{courant2024methods}, Eq.~\eqref{eq:tdepBoltz} has an exact solution using coordinates $\mathbf{x}_n(\mathbf{k},t)$ that solve the initial value problem
\begin{equation}
    \label{eq:chareq}
    \partial_t \mathbf{x}_n(\mathbf{k},t) = \mathbf{v}_n(\mathbf{k},t), \quad \mathbf{x}_n(\mathbf{k},0) = 0\,,
\end{equation}
in terms of which the solution to the initial value problem for Eq.~\eqref{eq:tdepBoltz} reads
\begin{equation}
    \rho_n(\mathbf{x},\mathbf{k},t) = \rho_n[\mathbf{x}-\mathbf{x}_n(\mathbf{k},t),\mathbf{k},0]\,.
\end{equation}
To test this prediction numerically, we simulate the well-studied periodically-driven Rice-Mele Hamiltonian
\begin{align}
    \nonumber
    \hat{H}_0(t) = \sum_{j=-L/2}^{L/2-1}\,\Big[&J(t)\,\hat c^\dagger_{j,A}\hat c^{}_{j,B}+\,\hat c^\dagger_{j,A}\hat c^{}_{j+1,B}+\text{H.c.}\Big] \\&+\Delta(t)\Big(\hat{c}_{j,A}^\dagger\hat{c}^{}_{j,A}-\hat{c}_{j,B}^\dagger\hat{c}^{}_{j,B}\Big)\,,
    \label{eq:tdephamchoice}
\end{align}
where $J(t) = m +\cos{(2 \pi t/\tau)}$ and $\Delta(t) =\sin{(2\pi t/\tau)}$, and the parameters $m$ and $\tau$ will be varied. The Bloch Hamiltonian is given by
\begin{equation}
    h(k,t) = \begin{pmatrix} \Delta(t) & J(t) + e^{ik} \\ J(t) + e^{-ik} & -\Delta(t)
    \end{pmatrix} \, ,
\end{equation}
and the instantaneous band structure is $E_{\pm}(k,t) =\pm\sqrt{\Delta(t)^2 + J(t)^2 + 2J(t)\cos{k}+1}$. For parameter values $0 < m <2$, this Hamiltonian has unit winding number in the $J-\Delta$ plane about the gap-closing point $\Delta=J-1=0$, implying Thouless pumping of one topologically quantized unit of charge per cycle in the adiabatic limit $\tau \to \infty$ at half filling and zero temperature~\cite{thouless1983quantization}. In practice, $\beta, \tau < \infty$ and perfect topological quantization is lost~\cite{Citro_2023}.

As in the previous section, the Heisenberg-picture charge density operator $\hat{n}_j = \hat{c}_{j,A}^\dagger\hat{c}^{}_{j,A}+\hat{c}_{j,B}^\dagger\hat{c}^{}_{j,B}$ satisfies the microsopic continuity equation in Eq.~\eqref{eq:continuity} for all times, where now the current density operator $\hat{j}_j = i(\hat{c}_{j-1,A}^\dagger\hat{c}^{}_{j,B}-\hat{c}_{j,B}^\dagger\hat{c}^{}_{j-1,A})$. Thus the total charge current $\hat{J}(t) = \sum_{j=-L/2}^{L/2-1} \hat{j}_j$ recovers Eq.~\eqref{eq:defJ} in the thermodynamic limit as $L \to \infty$.

Before addressing in detail the question of whether orbital Wigner functions can capture inherently topological effects such as Thouless pumping, we first investigate how far Eq.~\eqref{eq:tdepBoltz} captures the nonequilibrium dynamics of charge under time evolution by the Hamiltonian in Eq.~\eqref{eq:tdephamchoice} away from half filling and at nonzero temperature. Initial conditions for the numerical simulations have the same form as in Sec. \ref{sec:loceq} for the initial Hamiltonian $\hat{H}_0(0)$, with $L=400$ sites; constant initial temperature $\beta=1$ and inhomogeneous initial chemical potential as given in Eq.~\eqref{eq:chargespot} with length scale $l=40$; and bulk and central chemical potentials set to $\mu_0=-2$ and $\mu_1=2$, respectively. We fix the simulation time to be $t=100$ and vary the period of oscillation of the Hamiltonian over two orders of magnitude, from $\tau = 5$ to $500$. To obtain these predictions, we (i) use a converged discretization of the time-evolution operator in the basis of single-particle Bloch states to obtain microscopic free-fermion predictions, (ii) integrate the characteristic equationsin Eq.~\eqref{eq:chareq} numerically to solve the time-dependent Boltzmann equation, and (iii) continue to solve the transport Eq.~\eqref{eqn:tdepWignerEoM} using the staggered leapfrog method~\cite{press2007numerical}, now including time dependence of the Bloch Hamiltonian at each time step. Note that these methods are all applicable to arbitrary time dependence beyond periodic driving.

We find excellent agreement with the time-dependent Boltzmann equation in Eq.~\eqref{eq:tdepBoltz} in the regime $\tau \gtrsim t$ but marked discrepancies in the regime $\tau \ll t$, see Fig.~\ref{fig:loceqtdep}. The time-dependent Boltzmann description appears to freeze as $\tau$ is decreased below the longest simulation time and yields an increasingly poor description of the dynamics as $\tau \to 0$. However, the time-dependent Wigner function dynamics in Eq.~\eqref{eqn:tdepWignerEoM} continues to perform remarkably well even for short oscillation periods, with negligible ($<0.1\%$) error in the predicted charge density for all drive periods simulated. 

\subsection{Thouless pumping}

\begin{figure}[t]
    \centering
\includegraphics[width=0.99\linewidth]{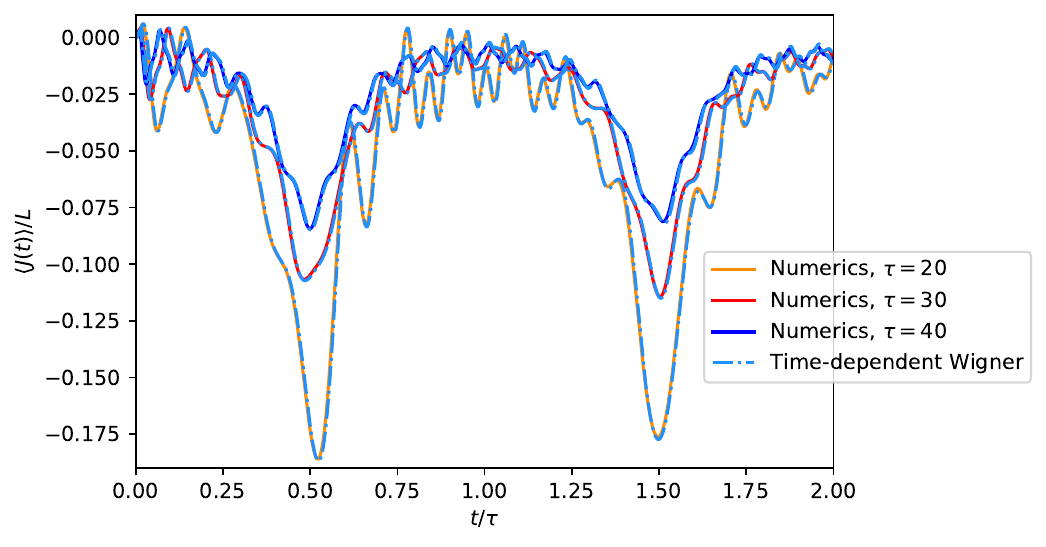}
\caption{Numerical time evolution of the current
density under translation invariant Thouless pumping at inverse temperature $\beta=100$. We simulate various time periods $\tau = 20,30,40$ and depict numerically obtained free-fermion predictions for the expected value of the current density operator $\langle \hat{j}_{j=0}(t)\rangle = \langle \hat{J}(t)\rangle/L$ at a given site (solid lines) against Wigner function predictions from Eq.~\eqref{eqn:tdepWignerEoM} over two time periods (blue dashed lines). By translation invariance, these two quantities should be exactly equal in the thermodynamic limit, which is consistent with the results shown.}
\label{fig:ThoulessCurr}
\end{figure}

We now consider the periodically driven Rice-Mele model given in Eq.~\eqref{eq:tdephamchoice} at half-filling $\mu=0$, while varying the inverse temperature $\beta$ and the driving period $\tau$. The canonical theoretical setting~\cite{thouless1983quantization} for Thouless pumping is recovered in the adiabatic, zero temperature limit $\tau,\beta \to \infty$. In practice, $\tau, \beta < \infty$ and there are two natural ways to probe charge pumping via variation of Hamiltonian parameters.

The first approach, which should be feasible in experiments capable of imaging correlation functions between neighboring unit cells, is to consider translation-invariant initial states and study the dynamics of the total charge current $\langle \hat{J}(t) \rangle$ directly. This approach is best suited for demonstrating theoretically that Eq.~\eqref{eqn:tdepWignerEoM} can capture Thouless pumping exactly in the adiabatic limit, see Eq.~\eqref{eq:JforThouless} below.

An alternative approach mimics experimental tests of Thouless pumping in ultracold atoms~\cite{nakajima2016topological} and consists of initializing a localized lump of charge, as in Eq.~\eqref{eq:chargespot}, before applying a slow variation of Hamiltonian parameters to induce pumping. We find that Eq.~\eqref{eqn:tdepWignerEoM} accurately captures such charge dynamics over multiple pumping cycles.

It is worth noting that the time-dependent Boltzmann equation in Eq.~\eqref{eq:tdepBoltz} is completely inadequate for describing Thouless pumping because it is invariant under spatial reflection and therefore predicts that for parity-even initial conditions, $\langle \hat{J}(t)\rangle=0$ for all time. Put differently, the nonzero current generated dynamically by Thouless pumping is incompatible with the assumption of instantaneous local equilibrium in Eq.~\eqref{eq:loceqansatz}, which can never yield a current-carrying state.

\subsubsection{Inducing a translation-invariant current}
\begin{figure*}[t]
    \centering
\includegraphics[width=0.99\linewidth]{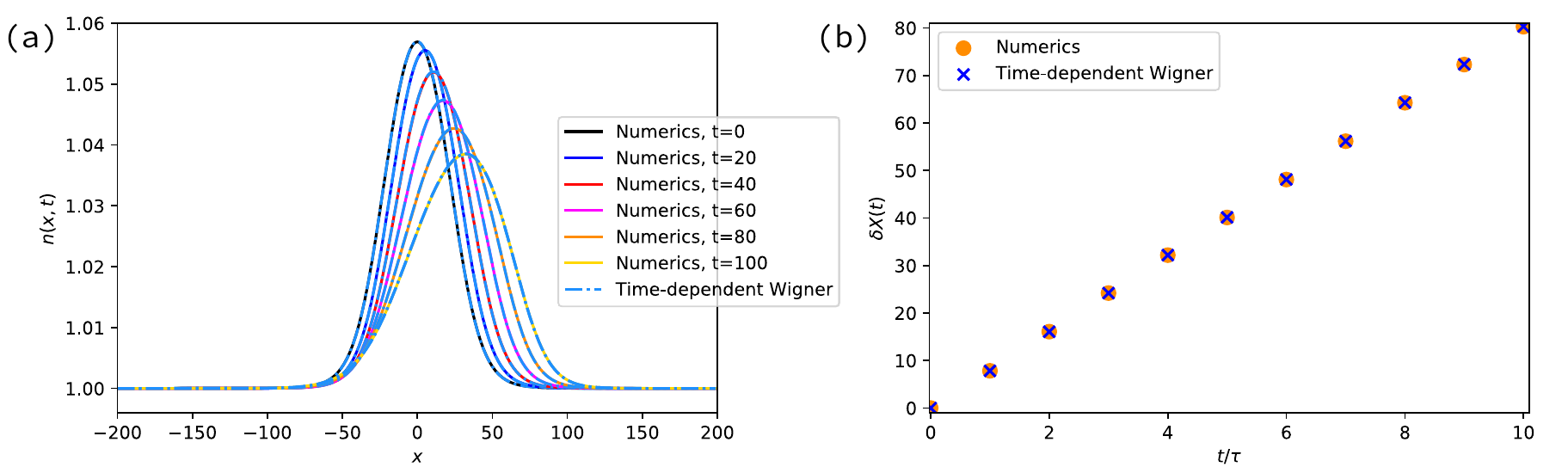}
\caption{(a) Numerical time evolution of an initially localized charge density under the periodic time evolution Eq.~\eqref{eq:tdephamchoice} for a time $t=100$ and driving period $\tau = 10$ compared with time-dependent Wigner function predictions (light blue dashed lines). The bulk temperature $\beta=10$ and the initial condition is as in Eq.~\eqref{eq:chargespot} with bulk chemical potential $\mu_0=0$, central chemical potential $\mu_1=1$ and characteristic width $l=80$. (b) The evolution of the first moment of charge density resulting from the dynamics on the left. The agreement between free-fermion numerics (orange dots) and Wigner function predictions (blue crosses) is excellent.}
    \label{fig:ThoulessTest}
\end{figure*}
We first consider the translation-invariant setting, in which a slow variation of Hamiltonian parameters induces a homogeneous, nonzero current in the system. Because of translation invariance, neither higher-order spatial derivative corrections to Eq.~\eqref{eqn:tdepWignerEoM} nor corrections to the initial local density approximation in Eq.~\eqref{eq:locthermwigner} can arise, so that the transport formalism given by Eqs.~\eqref{eq:locthermwigner} and \eqref{eqn:tdepWignerEoM} becomes exact in the thermodynamic limit, insofar as the underlying Hamiltonian Eq.~\eqref{eqn:Hamiltonian} is exact.

Concretely, we let the initial density matrix $\hat{\rho} \propto e^{-\beta \hat{H}_0(0)}$ and fix $\tau$. By translation invariance, the exact transport equation for the Wigner function takes the simple form
\begin{equation}
    \partial_t w(k,t) = -i\big[h(k,t),w(k,t)\big] \, ,
\end{equation}
where $w(x,k,t) = w(k,t)$ and the initial condition
\begin{equation}
    \label{eq:initW}
    w(k,0) = \frac{1}{1+e^{\beta h(k,0)}}\,.
\end{equation}
is also exact. Defining the single-particle propagator as the time-ordered exponential $
U(k,t) = \mathcal{T}\exp{\left[-i\int_{0}^t dt' h(k,t')\right]}$, it follows that
\begin{equation}
    w(k,t) = U(k,t)\,w(k,0)\,U^\dagger(k,t)\,.
\end{equation}
We then expect by the identity in Eq.~\eqref{eq:exactQWig} that
\begin{equation}
    \label{eq:JforThouless}
    \langle \hat{J}(t) \rangle/L \sim  \int_{\mathrm{BZ}} \frac{dk}{V_{\mathrm{BZ}}} \sum_{\alpha,\beta=1}^M \partial_{k} h_{\alpha \beta}(k)\, w_{\beta \alpha}(k,t)
\end{equation}
holds exactly in the thermodynamic limit, assuming that the current operator is given by Eq.~\eqref{eq:defJ} (the extension to off-diagonal current operators follows from Appendix \ref{appendix:offdiag}). Thus, Eq.~\eqref{eqn:tdepWignerEoM} should accurately capture Thouless pumping in finite, translation-invariant systems for all temperatures $\beta$ and driving periods $\tau$, up to finite-size corrections. Although it is arguably redundant, we verify this numerically in Fig.~\ref{fig:ThoulessCurr} by comparing the expectation value of the microscopic current density operator at a single site $\langle \hat{j}_{j=0}(t) \rangle$ against numerical integration of Eq.~\eqref{eq:JforThouless}, for $L=400$ sites, inverse temperature $\beta=100$ and various finite driving periods. Since the resulting dynamics is rather intricate, we restrict the simulation time to two driving periods for clarity and the agreement is excellent as expected.

\subsubsection{Pumping an inhomogeneous charge density}
We next consider pumping an initial, localized charge density in local equilibrium, which can be seen as a hydrodynamic version of the initial conditions realized experimentally in Ref. \cite{nakajima2016topological}. We set the bulk temperature $\beta=10$ to be relatively low and consider an inhomogeneous chemical potential as in Eq.~\eqref{eq:chargespot} with length scale $l = 80$. We set the bulk chemical potential to half-filling $\mu_0 = 0$ and the central chemical potential to $\mu_1 = 1$. We let the driving period $\tau = 10$ and simulate times up to $t=100$, i.e., 10 driving periods. This should be far from adiabaticity and thus provide an especially stringent test of Eq.~\eqref{eqn:tdepWignerEoM}. To highlight the physics of charge pumping, we further compute the evolution of the first moment of the charge density
\begin{equation}
    \delta X(t) = \sum_{x =-L/2}^{L/2-1} x \,\big[n(x,t)-n(x,0)\big]
\end{equation}
as a function of time, which grows linearly at a rate set by the initial conditions. Once again the agreement between theory and numerics is excellent, see Fig.~\ref{fig:ThoulessTest}.
\subsection{Charge transport from instantaneously flat bands}
As a final example, we consider the propagation of charge in stroboscopically driven SSH chains. This example is particularly striking because the group velocity of each quasiparticle vanishes instantaneously, but charge nevertheless spreads ballistically. Specifically, we consider evolution under the time-periodic Hamiltonian
\begin{equation}
\hat{H}_0(t) = \begin{cases} \hat{H}_{\mathrm{trivial}}, & n\tau \leq t < (n+\frac{1}{2})\tau, \\
\hat{H}_{\mathrm{topological}}, & (n+\frac{1}{2})\tau \leq t < (n+1)\tau, 
\label{eq:strobedrive}
\end{cases}
\end{equation}
for $n=0,1,2,\ldots$, where
\begin{align}
\nonumber 
\hat{H}_{\mathrm{trivial}} &=  \sum_{j=-L/2}^{L/2-1} \hat{c}^\dagger_{j,B} \hat{c}^{}_{j,A} + \mathrm{H.c.}\,, \\
\hat{H}_{\mathrm{topological}} &= \sum_{j=-L/2}^{L/2-1} \hat{c}^\dagger_{j+1,A} \hat{c}^{}_{j,B} + \mathrm{H.c.}\,,
\end{align}
and $\tau$ denotes the driving period. This model alternates stroboscopically between the two trivial and topological dimerized limits of the usual SSH model~\cite{su1980soliton}. (We note that various versions of this driven system have arisen before in the context of time-crystal physics~\cite{VedikaFloquet,CurtFloquet,DrewFloquet,Iadecola}, though the latter is unrelated to our focus here.) According to the semiclassical theory of transport, both the Hamiltonians $\hat{H}_{\mathrm{trivial}}$ and $\hat{H}_{\mathrm{topological}}$ are insulating, with flat-band spectra $E_1(k) = -E_2(k) = 1$ for all $k$ and therefore vanishing group velocities in each band.

Thus the time-dependent Boltzmann Eq.~\eqref{eq:tdepBoltz}, which assumes instantaneous local equilibrium, would predict no dynamics at all. However, a local equilibrium state with respect to $\hat{H}_{\mathrm{trivial}}$ will be far from equilibrium with respect to $\hat{H}_{\mathrm{topological}}$ and the resulting quenchlike dynamics turns out to be sufficient to yield ballistic spreading of charge. (This possibility could also have been deduced from the presence of bandcurvature in the Floquet-Bloch spectrum for this model~\cite{Iadecola} in a Floquet-Boltzmann approach~\cite{Genske_2015,Esin}.)

\begin{figure}[t]
    \centering
\includegraphics[width=0.99\linewidth]{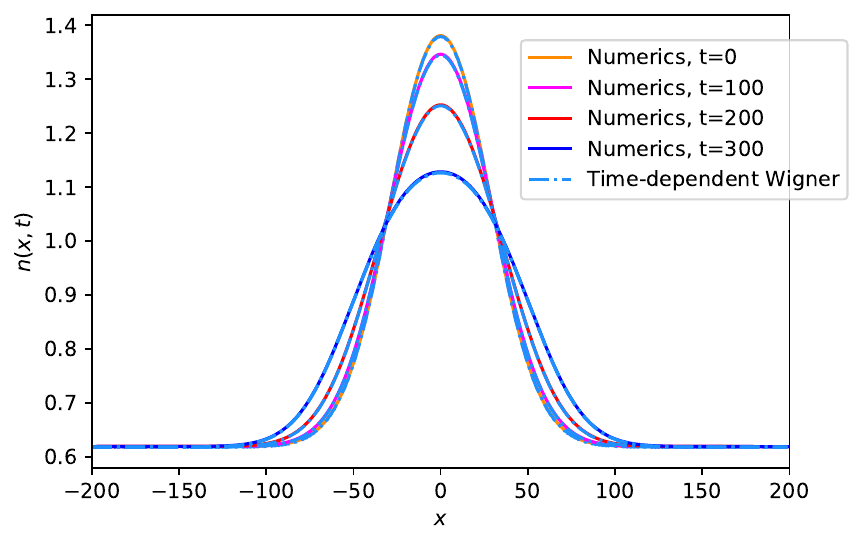}
\caption{Numerical time evolution of the charge density for the periodically driven SSH model Eq.~\eqref{eq:strobedrive} with driving period $\tau=5$, whose bands are always instantaneously flat. The drive nevertheless enables ballistic charge spreading that is accurately captured by the Wigner function prediction Eq.~\eqref{eqn:tdepWignerEoM}. The initial condition is as in Eq.~\eqref{eq:chargespot} with $\beta=1$, $\mu_0=-1$, $\mu_1=1$ and $l=40$ and agreement persists over 60 driving periods.}
\label{fig:FlatBands}
\end{figure}

The resulting dynamics is straightforwardly captured by Eq.~\eqref{eqn:tdepWignerEoM} with the stroboscopic Bloch Hamiltonian
\begin{equation}
    h(k,t) = \begin{cases} h_{\mathrm{trivial}}(k),& n\tau \leq t < (n+\frac{1}{2})\tau, \\
    h_{\mathrm{topological}}(k),& (n+\frac{1}{2})\tau \leq t < (n+1)\tau, 
\end{cases}
\end{equation}
where
\begin{align}
    \nonumber
    h_{\mathrm{trivial}}(k) &= \begin{pmatrix}
    0 & 1 \\ 1 & 0     
    \end{pmatrix}, \\
    h_{\mathrm{topological}}(k) &= \begin{pmatrix}
    0 & e^{-ik} \\ e^{ik} & 0     
    \end{pmatrix}. 
\end{align}
To test the orbital Wigner function predictions numerically, we prepare charge-spot initial conditions~\eqref{eq:chargespot} with $\mu_0 = -1$, $\mu_1 = 1$, bulk inverse temperature $\beta=1$ and length scale $l=40$ for the Hamiltonian $\hat{H}_{\mathrm{trivial}}$ on $L=400$ sites, before subjecting this to periodic driving as in Eq.~\eqref{eq:strobedrive} with driving period $\tau=5$.
See Fig.~\ref{fig:FlatBands} for a comparison between free-fermion numerics and the numerical solution to Eq.~\eqref{eqn:tdepWignerEoM}. We note that the discontinuous dynamics in Eq.~\eqref{eq:strobedrive} makes convergence of both the free-fermion numerics and Wigner function dynamics particularly challenging in this case, but once it is achieved the agreement is once again excellent.
\section{Discussion}
\label{sec:discussion}
We have presented an analytically transparent and numerically tractable theory of coherent electron transport in stationary and driven multiband (and more generally multiorbital) crystals. Our proposed formalism accurately captures far-from-equilibrium charge and current densities arising from non-uniform and quantum coherent initial conditions. We illustrated this point by simulating multiple distinct experimentally accessible Hamiltonians and various kinds of dynamics beyond standard semiclassical transport theory. While, in this work, we focus primarily on electric charge and current density, our approach allows for straightforward generalizations to spin, thermal, and thermoelectric transport through evaluating the corresponding charge and current operators from the orbital Wigner function. 

Although we have chosen to highlight the rich physics that can arise in multiband systems even in the absence of interactions and disorder, any practically useful transport theory must be able to model the dissipative hydrodynamic behavior that arises in realistic physical systems. On this front, authors of recent work on 1D systems have shown that the venerable quantum Boltzmann equation can yield remarkably quantitatively accurate results even in strong-coupling, long-time regimes~\cite{Bertini2015,bertini2016thermalization,zechmann2022tunable}, beyond the weak-coupling, short-time regime in which quantum kinetic theory might rigorously be expected to hold~\cite{erdHos2004quantum,SpohnBoltz}. 
We note that the collision terms appearing in the resulting equations would naturally be matrix valued~\cite{SpohnBoltz} and to that extent differ from standard formulations of the quantum Boltzmann equation with possibly nontrivial implications for hydrodynamics and transport. We leave a deeper investigation of this fascinating question for future work. 

A proper treatment of interactions will be necessary to obtain meaningful predictions for both linear and nonlinear conductivities and optical response from orbital Wigner functions. In this setting, it would be interesting to understand how far systematic derivations of the collision integral, along the lines sketched above, can capture nontrivial corrections to semiclassical transport theory. For example, the inclusion of a real-space shift during scattering events~\cite{Chang_1996, Xiao2019} was found to be necessary to yield a collision integral consistent with quantum kinetic theory at nonlinear order in the applied electric field~\cite{Gao2019, Kaplan2023}. Similarly, off-diagonal contributions to the position operator~\cite{Pedersen, Sandu, Iba_ez_Azpiroz_2022} are commonly neglected in tight-binding descriptions of condensed-matter systems but can nevertheless play an important role in nonlinear optical response. Such corrections could in principle be captured in our approach by a suitable modification of the two-body operators corresponding to the center of charge and the electrostatic potential energy, as discussed in Appendix \ref{appendix:offdiag}.

Thus orbital Wigner functions could provide a microscopic route toward deriving nonlinear response functions in settings where quantum coherence is important. Indeed, the fact that orbital Wigner functions correctly account for the plethora of quantum coherent effects studied in this paper strongly suggests that they can accurately capture other known topological and geometrical transport phenomena that arise in multiband systems, such as the Berry phase and the anomalous velocity~\cite{Xiao2010}. As an illustrative example, our approach effortlessly handles band crossings and near-degenerate bands, which can be challenging to treat using standard Boltzmann-type techniques. 

While we expect that our formalism encompasses these well-known physical phenomena, we emphasize that it also has the capacity to predict new ones. Recent results on Floquet theory \cite{Schindler2025}, nonlinear optical responses \cite{Ahn2022, Bouhon2023, Kaplan2023, Avdoshkin2024, Mitscherling2024, Jankowski2024, Jankowski2025}, and light-matter coupling \cite{Holder2020, Topp2021} raise the possibility that new and possibly unforeseen geometric invariants could play an essential role in nonequilibrium transport in both Hamiltonian and driven systems. The systematic approach pursued in these works and in the current paper suggest one potential strategy for identifying the most promising lattice systems and initial conditions for unconventional quantum-coherent transport to arise, for example, in flat-band systems \cite{Leykam2018, Poblete2021, Mitscherling2022, Danieli2024}. Looking beyond solid-state platforms, the numerical accuracy promised by our approach makes it ideally suited for a direct quantitative comparison with analog quantum simulators~\cite{Brown_2019,Ozawa2019, Citro_2023, Halimeh2025}, with the potential to both reveal shortcomings in their theoretical description and point to new physics. 

More generally, Wigner-function-based approaches have recently been applied successfully in contexts ranging from thermal transport~\cite{Simoncelli2019, Simoncelli2022} to bosonic systems~\cite{mangeolle2024quantumkineticequationthermal} and the bosonization of Fermi surfaces~\cite{Delacretaz2022,Park2024}, and it might be fruitful to connect these proposals to the approach taken here. While we do not expect the full gamut of possibilities for electron transport in crystals to succumb to the formalism outlined in this paper, a broader lesson from our work is that most conventional theories of transport are inadequate for confronting the sheer variety of large-scale dynamics that can arise even in minimal tight-binding models. Attaining quantitative accuracy in this simplest of settings, as we have sought to do here, strikes us as a prerequisite for developing a theory of transport in solids that is equipped to address the many long-standing challenges in this field.

\begin{acknowledgments}
We thank Alexander Avdoshkin, Pieter W. Claeys, Frank Schindler, Tha\'is V. Trevisan, and Christopher W. W\"achtler for stimulating discussions, Nick G. Jones and Ruben Verresen for collaboration on related work; and David Xiao for helpful guidance on transition to PyTorch.
J.M. was supported by the German National Academy of Sciences Leopoldina through Grant No. LPDS 2022-06 and, in part, by the Deutsche Forschungsgemeinschaft (DFG) under Grant cluster of excellence ct.qmat (EXC 2147, Project No. 390858490) and the Emmy Noether Program (Project No. 568440758). Work by D.S.B. and J.E.M. was supported by the MURI project TOPFORCE of the Air Force Office of Scientific Research under Grant No. FA9550-22-1-027.
\end{acknowledgments}

\section*{Data Availability}

The source code for a GPU implementation of the Wigner formalism is publicly available~\cite{dataAvailability}. Requests for further information or data should be sent to the authors.

\appendix

\section{Fourier transform conventions}
\label{appendix:FourierConventions}
We define the Fourier transform of a function $f:\Lambda \to \mathbb{C}$ by
\begin{equation}
f(\mathbf{k}) = \sum_{j \in 
\Lambda} f(\mathbf{R}_j) e^{-i\mathbf{k}\cdot \mathbf{R}_j} \, ,
\end{equation}
with inverse
\begin{equation}
f(\mathbf{R}_j) = \int_{\mathrm{BZ}} \frac{d \mathbf{k}}{V_{\mathrm{BZ}}} \, e^{i\mathbf{k}\cdot \mathbf{R}_j} f(\mathbf{k}),
\end{equation}
where $\mathrm{BZ}$ denotes the first Brillouin zone and $V_{\mathrm{BZ}}$ its volume. Thus the (dual-lattice-periodic extension of the) Dirac delta function reads 
\begin{equation}
\delta(\mathbf{k}) = \frac{1}{V_{\mathrm{BZ}}} \sum_{j \in \Lambda} e^{-i\mathbf{k}\cdot \bR_j} \, ,
\end{equation}
while the real-space Kronecker delta can be written as
\begin{equation}
\delta_{\mathbf{R}_j0} = \int_{\mathrm{BZ}} \frac{d\mathbf{k}}{V_{\mathrm{BZ}}} \, e^{i \mathbf{k}\cdot \mathbf{R}_j}.
\end{equation}
It will be useful to define the {\it Brillouin delta function}
\begin{equation}
\delta_{\mathrm{B}}(\bx) = \int_{\mathrm{BZ}} \frac{d\mathbf{k}}{V_{\mathrm{BZ}}} \, e^{i \mathbf{k}\cdot \bx}
\end{equation}
for general incommensurate positions $\bx$. Recall that this equals unity when $\mathbf{x}=\mathbf{0}$, vanishes for $\mathbf{x} \in \Lambda \setminus \mathbf{0}$, and resembles a product of sinc functions for general incommensurate $\mathbf{x}$. 

We define momentum-space fermion operators by
\begin{equation}
\hat{c}_{\alpha}(\mathbf{k}) = \sum_{j \in \Lambda} e^{-i\mathbf{k}\cdot \mathbf{R}_{j\alpha}} \hat{c}_{j\alpha},
\end{equation}
which are easily verified to satisfy canonical anticommutation relations
\begin{equation}
\{ \hat{c}^\dagger_{\alpha}(\bk),\hat{c}_{\alpha'}(\bk')\} = V_{\mathrm{BZ}}\delta(\mathbf{k}-\mathbf{k}')\delta_{\alpha \alpha'}
\end{equation}
and the inversion formula
\begin{equation}
\hat{c}_{j\alpha}= \int_{\mathrm{BZ}} \frac{d\mathbf{k}}{V_{\mathrm{BZ}}} e^{i\mathbf{k}\cdot \mathbf{R}_{j\alpha}} \hat{c}_{\alpha}(\bk).
\end{equation}

\section{Higher-order derivative corrections}
\label{appendix:HigherOrder}
We now present the full set of derivative corrections to the transport equation Eq.~\eqref{eqn:EoM} (i.e. due to evolution under $\hat{H}_0$ alone). Continuing the Taylor expansion of Eq.~\eqref{eqn:EoMSpatial} to higher orders in $\partial_{\mathbf{x}}$ yields the expression
\begin{align}
    \partial_t \hat{W}_{\alpha \beta}&(\bx,\bk,t) = \sum_{m=0}^\infty \frac{i^{m+1}}{m!2^m} \sum_{\vec{i} \in \{1,2,\ldots,d\}^m} \nonumber\\&\times\sum_{\gamma=1}^M \partial^{(m)}_{k_{\vec{i}}} h_{\gamma\alpha}(\bk)\partial^{(m)}_{x_{\vec{i}}} \hat{W}_{\gamma \beta}(\bx,\bk,t) \nonumber \\ &\,\,\,\,\,\,\,\,\,\,- (-1)^m \partial^{(m)}_{k_{\vec{i}}} h_{\beta \gamma}(\bk) \partial^{(m)}_{x_{\vec{i}}} \hat{W}_{\alpha \gamma}(\bx,\bk,t) \, ,
\end{align}
where in each term we use the notation $\partial^{(m)}_{k_{\vec{i}}} := \partial_{k_{i_1}}\partial_{k_{i_2}}\ldots\partial_{k_{i_m}}$ and $\partial^{(m)}_{x_{\vec{i}}} := \partial_{x_{i_1}}\partial_{x_{i_2}}\ldots\partial_{x_{i_m}}$ and $\vec{i} = (i_1,i_2,\ldots,i_m)$ denotes the corresponding multi-index. It follows by linearity of this equation in $\hat{W}$ that the time evolution of the orbital Wigner function Eq.~\eqref{eqn:OrbitalWignerFunction} can be written as
\begin{align}
    \partial_t w_{\alpha \beta}&(\bx,\bk,t) = \sum_{m=0}^\infty \frac{i^{m+1}}{m!2^m} \sum_{\vec{i} \in \{1,2,\ldots,d\}^m} \nonumber \\ &\times\sum_{\gamma=1}^M \partial_{x_{\vec{i}}}^{(m)} w_{\alpha \gamma}(\bx,\bk,t)\, \partial^{(m)}_{k_{\vec{i}}} h_{\gamma\beta}(\bk) \nonumber \\ &\,\,\,\,\,\,\,\,\,\,- (-1)^m \partial^{(m)}_{k_{\vec{i}}} h_{\alpha \gamma}(\bk)\, \partial_{x_{\vec{i}}}^{(m)} w_{\gamma \beta}(\bx,\bk,t)\,.
\end{align}
However, the summation over internal indices is just ordinary matrix multiplication, and suppressing matrix indices we can write this more simply as
\begin{align}
\nonumber
&\partial_t w(\bx,\bk,t) \\&= -i\sum_{m=0}^\infty \frac{(-1)^m}{(2m)!2^{2m}} \sum_{\vec{i} \in \{1,2,\ldots,d\}^m}  \!\!\!\!\!\!\!\!\![\partial^{(2m)}_{k_{\vec{i}}} h(\bk),\partial^{(2m)}_{x_{\vec{i}}} w(\bx,\bk,t)] \nonumber \\
&- \!\sum_{m=0}^\infty \frac{(-1)^m}{(2m+1)!2^{2m+1}} \nonumber \\&\hspace{1cm}\times\!\!\sum_{\vec{i} \in \{1,2,\ldots,d\}^m} \!\!\!\!\!\!\!\!\! \{\partial^{(2m+1)}_{k_{\vec{i}}} h(\bk),\partial^{(2m+1)}_{x_{\vec{i}}} w(\bx,\bk,t)\}.
\label{eq:FullEoM}
\end{align}
We note that whenever $h(\bk)$ is differentiable, the linear-order term in spatial derivatives in Eq.~\eqref{eq:FullEoM} only vanishes if all other higher-order terms vanish. Given that ballistic transport, when present, dominates the dynamics, the above formula shows that finite-range, non-interacting, translation-invariant Hamiltonians are either trivially localized or exhibit ballistic transport.

For convenience, we record the first few terms in this expansion:
\begin{align}
    \nonumber 
    &\partial_t w(\bx,\bk,t) = -i[h(\bk),w(\bx,\bk,t)] \\ &- \frac{1}{2}\sum_{a=1}^d \{\partial_{k_a} h(\bk),\partial_{x_a} w(\bx,\bk,t)\} \nonumber \\&+ \frac{i}{8} \sum_{a,b=1}^d [\partial_{k_a}\partial_{k_b} h(\bk),\partial_{x_a}\partial_{x_b} w(\bx,\bk,t)] \nonumber \\
    &+ \frac{1}{48} \sum_{a,b,c=1}^d \{\partial_{k_a} \partial_{k_b} \partial_{k_c} h(\bk),\partial_{x_a} \partial_{x_b} \partial_{x_c} w(\bx,\bk,t)\} \nonumber \\&+ O(\partial_{\bx}^4) \, .
\end{align}
For a single band, the even-order terms vanish and this expression recovers known results on dispersive corrections to free-fermion dynamics~\cite{fagotti2017higher,essler2023short}. For multiple bands, the second-order term becomes the leading dispersive correction. Note that all higher-order corrections to Eq.~\eqref{eq:FullEoM}, including this second-order term, are time-reversal invariant and therefore do not describe irreversible effects such as diffusion, which is consistent with the microscopic reversibility of the underlying Hamiltonian dynamics.

\section{Extension to static electric fields}
\label{appendix:extfield}
\subsection{Equation of motion and Bloch oscillations}
An important feature of any transport theory is its ability to make predictions in the presence of external electric and magnetic fields. Since this paper is dedicated to the nontrivial quantum effects that arise even in the absence of such fields, we will largely defer a detailed discussion of electric fields (and all discussion of magnetic fields) to future work.\\
\indent However, for completeness we derive a ballistic-scale equation of motion in the presence of an external electrostatic potential
\begin{equation}
\label{eq:defV}
    \hat{V} = \sum_{j \in \Lambda} \sum_{\alpha=1}^M V(\mathbf{R}_{j\alpha})\, \hat{c}^\dagger_{j \alpha} \hat{c}^{}_{j\alpha}\,,
\end{equation}
so that the total Hamiltonian becomes
\begin{equation}
    \hat{H} = \hat{H}_0 + \hat{V}.
\end{equation}
(Note that we have neglected off-diagonal contributions to $\hat{V}$~\cite{Pedersen,Sandu,Iba_ez_Azpiroz_2022}, which limits our accuracy to linear order in applied electric fields. However, such corrections could in principle be included in our approach, see Appendix \ref{appendix:offdiag}, and would modify Eq.~\eqref{eqn:FieldEoM} accordingly.) In position space and letting $\overline{\mathbf{R}}^{\alpha \beta}_{jj'} =(\mathbf{R}_{j\alpha}+\mathbf{R}_{j'\beta})/2$, the orbital Wigner operator can be expressed in terms of the Brillouin delta function $\delta_{\mathrm{B}}(\bx)$ (see Appendix \ref{appendix:FourierConventions}) as
\begin{equation}
    \label{eq:posspaceW}
    \hat{W}_{\alpha \beta}(\mathbf{x},\mathbf{k}) = \sum_{j,j'\in\Lambda} \delta_{\mathrm{B}}\Big(\mathbf{x}-\overline{\mathbf{R}}^{\alpha \beta}_{jj'}\Big) \,e^{i\mathbf{k}\cdot (\bR_{j\alpha}-\bR_{j'\beta}) }\, \hat{c}^\dagger_{j\alpha}\hat{c}^{}_{j'\beta}\,.
\end{equation}
Since $\hat{V}$ is diagonal in the Wannier basis by assumption, we have $i[\hat{V},\hat{c}_{j\alpha}] = - iV(\mathbf{R}_{j\alpha})\, \hat{c}_{j\alpha}$, and it follows by Eq.~\eqref{eq:posspaceW} that
\begin{align}
    &i[\hat{V},\hat{W}_{\alpha \beta}(\mathbf{x},\mathbf{k})] \nonumber \\ &= i \sum_{j,j'\in \Lambda}\delta_{\mathrm{B}}(\mathbf{x}-\overline{\mathbf{R}}^{\alpha \beta}_{jj'}) \big(V(\mathbf{R}_{j\alpha}) - V(\mathbf{R}_{j'\beta})\big) \nonumber \\ &\hspace{30mm}\times e^{i\mathbf{k}\cdot (\bR_{j\alpha}-\bR_{j'\beta}) } \hat{c}^\dagger_{j\alpha}\hat{c}^{}_{j'\beta}\,.
\end{align}
Under the summation, we can Taylor expand
\begin{equation}
    V(\mathbf{R}_{j\alpha}) - V(\mathbf{R}_{j'\beta}) = (\bR_{j\alpha}-\bR_{j'\beta}) \cdot \partial_{\mathbf{x}} V(\mathbf{x}) + O(\partial^2_{\mathbf{x}} V)
\end{equation}
to yield
\begin{equation}
    i[\hat{V},\hat{W}_{\alpha \beta}(\mathbf{x},\mathbf{k})] =  \partial_{\mathbf{x}} V(\mathbf{x}) \cdot \partial_{\mathbf{k}} \hat{W}_{\alpha \beta}(\mathbf{x},\mathbf{k}) + O(\partial^2_{\mathbf{x}} V)\,.
\end{equation}
For a generic nonuniform electrostatic potential $V(\mathbf{R}) = -e \phi(\mathbf{R})$, following the manipulations of the previous Appendix \ref{appendix:HigherOrder} then yields the ballistic-scale transport equation  
\begin{align}
    \partial_t w(\bx,\bk,t) &= -i[h(\bk),w(\bx,\bk,t)]\nonumber\\ 
    &-\frac{1}{2}\sum_{a=1}^d\{\partial_{k_a}h(\bk),\partial_{x_a} w(\bx,\bk,t) \} \nonumber \\&-  e\,\partial_{\bx} \phi(\mathbf{x}) \cdot \partial_{\mathbf{k}} w(\bx,\bk,t)\,
    \label{eqn:FieldEoM}
\end{align}
at leading order in spatial derivatives.\\
\indent For a single band for which the Wigner and Boltzmann formalisms coincide, this recovers the standard Boltzmann forcing term~\cite{ashcroft1976solid} as expected. Using this method, one can in principle compute higher-order corrections to the transport equation in momentum derivatives of $w$ but the lack of spatial localization of the Brillouin delta function $\delta_{\mathrm{B}}(\mathbf{x})$ means that the resulting expressions are not very illuminating. We note that Eq.~\eqref{eqn:FieldEoM} generalizes straightforwardly to time-dependent Hamiltonians and potentials, provided the latter are accompanied by negligible magnetic fields.
\\ \indent
In the special case of a spatially uniform electrostatic field $\phi(\bx) = - \mathbf{E} \cdot \bx$ (and treating the position operator as diagonal), such higher-order corrections in $\partial_{\bk}$ vanish and Eq.~\eqref{eqn:FieldEoM} admits a simplification analogous to the emergence of Bloch oscillations in ordinary semiclassical transport theory. To see this, suppose that the electrostatic field is uniform and define the accelerating-frame Wigner function $u(\bx,\bk,t)$ via the relation
\begin{equation}
w(\bx,\bk,t) = u(\bx,\bk + e\mathbf{E}t,t)
\end{equation}
at each time $t$. Then $u$ satisfies the force-free but time-dependent system of PDEs
\begin{align}
    \partial_t u(\bx,\bk,t) &= -i[h(\bk-e\mathbf{E}t),u(\bx,\bk,t)]\nonumber\\ 
    &-\frac{1}{2}\sum_{a=1}^d\{\partial_{k_a}h(\bk-e\mathbf{E}t),\partial_{x_a} u(\bx,\bk,t) \},
    \label{eqn:AccelFieldEoM}
\end{align}
which is a special case of Eq.~\eqref{eqn:tdepWignerEoM}. Thus, just as in the usual semiclassical theory of Bloch oscillations~\cite{ashcroft1976solid}, it is always possible to eliminate a uniform, time-independent electrostatic field in Eq.~\eqref{eqn:FieldEoM} by passing to a uniformly accelerating frame of reference, at the price of making the effective Bloch Hamiltonian time dependent. 
\\ \indent 
The equation of motion for the orbital Wigner function is given in Eq.~\eqref{eqn:FieldEoM}. We now derive the equation of motion for the band Wigner function $\tilde{w}(\bx,\bk,t) = U^\dagger(\bk)\, w (\bx,\bk,t)\,U(\bk)$ introduced in Eq.~\eqref{eqn:bandWigner} for a spatially uniform electrostatic field $\phi(\bx) = - \mathbf{E} \cdot \bx$. By insertion of the definition and using unitarity, we obtain
\begin{align}
    &\partial_t \tilde w(\bx,\bk,t) =-i[\tilde h(\bk),\tilde w(\bx,\bk,t)]\nonumber\\&\,\,\,\,\,-\frac{1}{2}\sum_{a=1}^d\{U^\dagger(\bk)\partial_{k_a}h(\bk)U(\bk),\partial_{x_a}\tilde w(\bx,\bk,t)\} \nonumber \\&\,\,\,\,\,+e\,\mathbf{E} \cdot U^\dagger(\bk)\partial_{\mathbf{k}} \big(U(\bk)\tilde w(\bx,\bk,t)U^\dagger(\bk)\big)U(\bk) \, ,
    \label{eqn:DerivationEFieldBandBasis}
\end{align}
where we introduced the diagonalized Bloch Hamiltonian $\tilde{h}(\bk) = U^\dagger(\bk)\, h (\bk)\,U(\bk)$. The generalized velocity decomposes as $U^\dagger(\bk) \,\big(\partial_{k_a} h(\bk)\big)\, U(\bk) = \partial_{k_a} \tilde{h}(\bk) + i\big[\mathcal{A}_a(\bk),\tilde{h}(\bk)\big]$ as in Eq.~\eqref{eq:BerryConnAppears}, where the Berry connection $\mathcal{A}(\bk) = -i U^\dagger(\bk)\, \partial_{\bk} U(\bk)$ with components $\mathcal{A}_a(\bk)$ arises by momentum dependence of the Bloch states. Applying the same decomposition to the last term of Eq.~\eqref{eqn:DerivationEFieldBandBasis} yields the ballistic-scale transport equation in the band basis
\begin{align}
    &\partial_t \tilde w(\bx,\bk,t) =-i[\tilde h(\bk)-e\,\mathbf{E}\cdot\mathcal{A}(\bk),\tilde w(\bx,\bk,t)]\nonumber\\&-\frac{1}{2}\sum_{a=1}^d\{\partial_{k_a} \tilde{h}(\bk) + i\big[\mathcal{A}_a(\bk),\tilde{h}(\bk)\big],\partial_{x_a}\tilde w(\bx,\bk,t)\} \nonumber\\&+e\,\mathbf{E} \cdot \partial_{\mathbf{k}}\tilde w(\bx,\bk,t) \, .
\end{align}
We note that this equation of motion yields two terms proportional to the uniform electrostatic field, which combined ensure invariance under gauge transformations of the Bloch states. The first term on the right-hand side also includes (gauge-dependent) diagonal matrix elements of $\mathbf{E}\cdot \mathcal{A}(\bk)$, which have a non-trivial effect whenever the matrix $\tilde{w}$ is non-diagonal, as arises generically in the quantum coherent settings considered in the main text.

\subsection{Bloch oscillations at higher orders in spatial derivatives}
In fact, this standard semiclassical prescription for treating Bloch oscillations~\cite{ashcroft1976solid} continues to hold at higher order in spatial derivatives. The first step in showing this is to note that in the presence of a static electric field, the exact Heisenberg equation of motion for the Wigner operator Eq.~\eqref{eqn:EoMSpatial} is modified to 
\begin{align}
    &\partial_t \hat W_{\alpha\beta}(\bx,\bk,t) - e \mathbf{E} \cdot \partial_{\bk} \hat{W}_{\alpha\beta}(\bx,\bk,t)  \nonumber \\ &= -i\sum_{\gamma=1}^M \sum_{\br \in \Lambda}\,\Big(t^{\gamma\alpha}(\br)\,e^{-i\bk\cdot \br_{\gamma\alpha}}\,\hat W_{\gamma\beta}\big(\bx+\br_{\gamma\alpha}/2,\bk,t\big)\nonumber\\&\hspace{10mm}-t^{\beta\gamma}(\br)\,e^{-i\bk\cdot \br_{\beta\gamma}}\,\hat W_{\alpha\gamma}\big(\bx-\br_{\beta\gamma}/2,\bk,t\big) \Big)\,.
     \label{eqn:EoMSpatialField}
\end{align}
We next define the accelerating-frame Wigner operator $\hat{U}_{\alpha\beta}(\bx,\bk,t)$ for all time via the relation
\begin{equation}
\hat{W}(\bx,\bk,t) = \hat{U}(\bx,\bk+e\mathbf{E}t,t).
\end{equation}
Substituting into Eq.~\eqref{eqn:EoMSpatialField} yields
\begin{align}
    &\partial_t \hat U_{\alpha\beta}(\bx,\bk,t) \nonumber \\ &= -i\sum_{\gamma=1}^M \sum_{\br \in \Lambda}\,\Big(t^{\gamma\alpha}(\br)\,e^{-i(\bk-e\mathbf{E}t)\cdot \br_{\gamma\alpha}}\,\hat U_{\gamma\beta}\big(\bx+\br_{\gamma\alpha}/2,\bk,t\big) \nonumber \\ &\,\,\,\,\,-t^{\beta\gamma}(\br)\,e^{-i(\bk-e\mathbf{E}t)\cdot \br_{\beta\gamma}}\,\hat U_{\alpha\gamma}\big(\bx-\br_{\beta\gamma}/2,\bk,t\big)\Big).
\end{align}
Taylor expanding as in the previous appendix then yields the derivative expansion
\begin{align}
    &\partial_t \hat{U}_{\alpha \beta}(\bx,\bk,t)  = \sum_{m=0}^\infty \frac{i^{m+1}}{m!2^m} \sum_{\vec{i} \in \{1,2,\ldots,d\}^m} \nonumber \\ & \hspace{8mm} \times\sum_{\gamma=1}^M \partial_{x_{\vec{i}}}^{(m)} \hat{U}_{\alpha \gamma}(\bx,\bk,t)\, \partial^{(m)}_{k_{\vec{i}}} h_{\gamma\beta}(\bk-e\mathbf{E}t) \nonumber \\ &\hspace{8mm}- (-1)^m \partial^{(m)}_{k_{\vec{i}}} h_{\alpha \gamma}(\bk-e\mathbf{E}t)\, \partial_{x_{\vec{i}}}^{(m)} \hat{U}_{\gamma \beta}(\bx,\bk,t)\,.
\end{align}
Finally, defining accelerating-frame Wigner functions $u_{\alpha\beta}(\bx,\bk,t) = \langle \hat{U}_{\beta\alpha}(\bx,\bk,t)\rangle$, we deduce the transport equation
\begin{align}\label{eqn:full_eom}
    \nonumber
    &\partial_t u(\bx,\bk,t) = \nonumber \\&-i\sum_{m=0}^\infty \frac{(-1)^m}{(2m)!2^{2m}} \nonumber \\ &\hspace{5mm}\times\!\!\!\!\sum_{\vec{i} \in \{1,2,\ldots,d\}^m}  [\partial^{(2m)}_{k_{\vec{i}}} h(\bk-e\mathbf{E}t),\partial^{(2m)}_{x_{\vec{i}}} u(\bx,\bk,t)] \nonumber \\
    &- \sum_{m=0}^\infty \frac{(-1)^m}{(2m+1)!2^{2m+1}} \nonumber \\&\hspace{5mm}\times\!\!\!\sum_{\vec{i} \in \{1,2,\ldots,d\}^m}  \{\partial^{(2m+1)}_{k_{\vec{i}}} h(\bk-e\mathbf{E}t),\partial^{(2m+1)}_{x_{\vec{i}}} u(\bx,\bk,t)\}\,.
\end{align}
Truncating this at leading order in spatial derivatives recovers Eq.~\eqref{eqn:AccelFieldEoM} above.
\section{Off-diagonal matrix elements of position, charge and related operators}
\label{appendix:offdiag}
\subsection{Overview}
In the main text, we studied fully general tight-binding Hamiltonians of the form
\begin{equation}
\hat{H}_0 = -\sum_{j,j' \in \Lambda} \sum_{\alpha,\beta=1}^Mt_{jj'}^{\alpha\beta} \hat{c}_{j\alpha}^\dagger \hat{c}_{j'\beta}
\end{equation}
and the exact dynamics of fermion bilinears arising from $\hat{H}_0$. We were particularly interested in charge transport.To model this, one needs a local definition of the charge density. While the total charge operator is defined unambiguously as $\hat{N} = \sum_{j\in\Lambda} \sum_{\alpha=1}^M \hat{n}_{j\alpha}$, the local charge density $\hat{n}_j$ is only defined up to a (discrete) total derivative. In more detail, any sensible definition of $\hat{n}_j$ should be
\begin{enumerate}
    \item localized to the cell $j$,
    \item covariant under translations, and
    \item satisfy the sum rule $\hat{N} = \sum_{j \in \Lambda} \hat{n}_j$.
\end{enumerate}
Clearly given any operator $\hat{o}_j$ that satisfies the first two conditions and $\sum_{j\in\Lambda} \hat{o}_j = 0$, we are free to redefine the charge density as $\hat{n}_j \mapsto \hat{n}_j + \hat{o}_j$. This ambiguity of hydrodynamic theory is well known~\cite{Fagotti_2020}.
\\ \indent
Above, we made the specific choice  
\begin{equation}
\label{eq:papercharge}
\hat{n}_j = \sum_{\alpha=1}^{M} \hat{n}_{j\alpha}.
\end{equation}
Physical assumptions that would naturally give rise to this definition are (i) perfect localization of the physical charge at Wannier centers $\mathbf{R}_{j\alpha}$ and (ii) no transitions between pairs of Wannier orbitals. Implicitly making these assumptions led us to the specific definitions 
\begin{equation}
\label{eq:papercurr}
\hat{\mathbf{J}} = \frac{d}{dt} \hat{\mathbf{X}}, \quad \hat{\mathbf{X}} = \sum_{j\in \Lambda}\sum_{\alpha=1}^M \mathbf{R}_{j\alpha} \hat{n}_{j\alpha}
\end{equation}
for the total current operator and
\begin{equation}
\label{eq:paperpot}
\hat{V} = \sum_{j \in \Lambda}\sum_{\alpha=1}^M V(\mathbf{R}_{j\alpha})\hat{n}_{j\alpha}
\end{equation}
for the two-body operator corresponding to an external electrostatic potential $V(\mathbf{x})$. 
\\
\indent Let us refer to this approximation [equivalently Eqs.~\eqref{eq:papercharge}-\eqref{eq:paperpot}] as the {\it perfect tight-binding approximation}. This appendix explains why the perfect tight-binding approximation can be relaxed within our formalism without affecting the validity of the transport equation Eq.~\eqref{eqn:EoM}. Even if we no longer assume that physical charge is perfectly localized at Wannier centers $\mathbf{R}_{j\alpha}$ and allow for interorbital transitions whenever Wannier orbitals overlap, Eq.~\eqref{eqn:EoM} should still yield reliable predictions for ballistic-scale charge and current dynamics.
\\
\indent We can show this systematically by starting from a microscopic definition of the charge density $\hat{n}(\mathbf{x})$, which then yields a specific microscopic expression for the two-body operator $\hat{V}$ corresponding to an external electrostatic potential $V(\mathbf{x})$, namely
\begin{equation}
\label{eq:micrV}
\hat{V} = \int d \mathbf{x} \, V(\mathbf{x})\hat{n}(\mathbf{x}) \,,
\end{equation}
which differs in general from Eq.~\eqref{eq:paperpot}. Similarly, starting from a spatially local and globally $U(1)$ symmetric microscopic model, the Heisenberg-picture  spatial charge density must always satisfy a continuity equation of the form
\begin{equation}
\partial_t \hat{n} + \partial_\mathbf{x} \cdot \hat{\mathbf{j}} = 0
\end{equation}
by Noether's theorem, which in turn fixes the microscopic definition of the total current operator $\hat{\mathbf{J}} = \int d \mathbf{x} \, \hat{\mathbf{j}}(\mathbf{x})$ to be
\begin{equation}
\label{eq:micrJ}
\hat{\mathbf{J}} = \frac{d}{dt} \hat{\mathbf{X}}, \quad  \hat{\mathbf{X}} = \int d\mathbf{x} \, \mathbf{x} \, \hat{n}(\mathbf{x}) \,.
\end{equation}
We now explain how orbital Wigner functions can be used to compute $\hat{n}(\bx), \, \hat{\mathbf{J}}$ and $\hat{V}$ in this more general setting.
\subsection{Restoring off-diagonal matrix elements}
As our microscopic starting point, we consider the continuum single-particle Hamiltonian
\begin{equation}
\hat{H} = \frac{\|\hat{\mathbf{p}}\|^2}{2m} + U(\hat{\bx})\,,
\end{equation}
where $U(\bx)$ combines ionic and self-consistent contributions to the effective one-particle potential and has the periodicity of a Bravais lattice $\Lambda \subset \mathbb{R}^d$. For simplicity, we consider spinless fermionic quantum fields $\hat{\psi}(\bx)$ (the generalization to nonzero pseudospin is straightforward) and an orthonormal basis of Wannier states $\varphi_{j\alpha}(\bx)$ with Wannier centers $
\mathbf{R}_{j\alpha} = \int d\bx \, |\varphi_{j\alpha}(\bx)|^2\bx$, where $j$ denotes the lattice index and $\alpha$ the basis as usual. Then expanding
\begin{equation}
\hat{\psi}(\bx) = \sum_{j\in \Lambda}\sum_{\alpha=1}^M \varphi_{j\alpha}(\bx) \hat{c}_{j\alpha},
\end{equation}
the second-quantized Hamiltonian can be written as
\begin{equation}
\hat{H}_0 = -\sum_{j,j' \in \Lambda} \sum_{\alpha,\beta=1}^Mt_{jj'}^{\alpha\beta} \hat{c}_{j\alpha}^\dagger \hat{c}_{j'\beta}\,,
\end{equation}
where $t_{jj'}^{\alpha \beta} = -\int d\bx \, \varphi_{j\alpha}^*(\bx) \left[-\frac{\hbar^2}{2m}\partial_\bx^2 + U(\bx)\right] \varphi_{j'\beta}(\bx)$.
\\
\indent In the main text, we defined lattice densities of global conserved fermion bilinears $\hat{Q}$ as local fermionic operators $\hat{q}_j$ such that
$\hat{Q} = \sum_{j \in \Lambda} \hat{q}_j$. However, for physical fermionic systems defined in the continuum, true charge densities $\hat{q}(\mathbf{x})$ are operators satisfying $
\hat{Q} = \int d\mathbf{x} \, \hat{q}(\mathbf{x})$. (As on the lattice, these are only unique up to a total derivative.)\\
For the system under consideration, the local charge density $\hat{n}(\bx) = \hat{\psi}^\dagger(\bx) \hat{\psi}(\bx)$ is given by
\begin{align}
\label{eq:micrchargedens}
\hat{n}(\bx) = \sum_{j,j'\in\Lambda}\sum_{\alpha,\beta=1}^M \varphi_{j\alpha}^*(\bx) \varphi_{j'\beta}(\bx) \hat{c}_{j\alpha}^\dagger \hat{c}_{j'\beta}.
\end{align}
Thus an electrostatic potential $V(\mathbf{x})$ defines the two-body operator
\begin{align}
\nonumber 
\hat{V} &= \sum_{j,j'\in\Lambda} \sum_{\alpha,\beta=1}^M V_{jj'}^{\alpha \beta}\hat{c}_{j\alpha}^\dagger \hat{c}_{j'\beta}, \\
V_{jj'}^{\alpha \beta} &= \int d\bx \, \varphi_{j\alpha}^*(\bx) \, V(\bx) \, \varphi_{j'\beta}(\bx) 
\end{align}
by Eq.~\eqref{eq:micrV} and the center-of-charge operator becomes
\begin{align}
\nonumber
\hat{\mathbf{X}} &= \sum_{j,j'\in\Lambda}\sum_{\alpha,\beta=1}^M \mathbf{X}_{jj'}^{\alpha\beta}\hat{c}_{j\alpha}^\dagger \hat{c}_{j'\beta}, \\
\mathbf{X}_{jj'}^{\alpha\beta} &= \int d \bx \, \varphi_{j\alpha}^*(\bx) \, \bx \,\varphi_{j'\beta}(\bx).
\end{align}
The total current operator is given by 
\begin{equation}
\hat{\mathbf{J}} = i[\hat{H}_0,\hat{\mathbf{X}}] = -i\sum_{j,j' \in \Lambda} \sum_{\alpha,\beta} [t,\mathbf{X}]^{\alpha \beta}_{jj'} \hat{c}_{j\alpha}^\dagger \hat{c}_{j'\beta}.
\end{equation}
\indent It is clear from the above discussion that the off-diagonal contributions to the charge density, current and potential energy operators can be highly nonlocal in the Wannier basis in general, and indeed this is expected if the Wannier functions are highly extended, as might arise for a topological band~\cite{Brouder_2007}. To compute such off-diagonal expectation values, we need a method for obtaining general fermion bilinears $\hat{c}^\dagger_{j\alpha} \hat{c}_{j'\beta}$ from the orbital Wigner function. This can be done using the identity
\begin{equation}
\label{eq:offdiagWigner}
\int_{\mathrm{BZ}} \frac{d\mathbf{k}}{V_{\mathrm{BZ}}} e^{i\bk\cdot(\bR_{j'\beta}-\bR_{j\alpha})}\hat{W}\Big(\overline{\bR}^{\alpha\beta}_{jj'},\bk\Big) = \hat{c}^\dagger_{j\alpha}  \hat{c}_{j'\beta}.
\end{equation}
Thus the complexity of computing the expectation value of a given off-diagonal operator depends on the range of off-diagonal contributions to $\hat{n}(\bx)$, $\hat{\mathbf{J}}$ and $\hat{\mathbf{V}}$ that are considered. Given the tradeoff between practicality and accuracy, this range is typically short~\cite{Iba_ez_Azpiroz_2022}, and this complexity is correspondingly low.
\\
\indent As an example, the microscopic charge density
\begin{align}
\nonumber
\hat{n}(\mathbf{x}) = \sum_{j,j'\in\Lambda} \sum_{\alpha,\beta = 1}^M \int_{\mathrm{BZ}} &\frac{d\mathbf{k}}{V_{\mathrm{BZ}}} e^{i\bk\cdot(\bR_{j'\beta}-\bR_{j\alpha})} \varphi_{j\alpha}^*(\bx) \varphi_{j'\beta}(\bx)\\
\label{eq:candidatemotivation}
&\times\hat{W}_{\alpha\beta}(\overline{\bR}^{\alpha\beta}_{jj'},\bk).
\end{align}
Evolving the Wigner function in time according to Eq.~\eqref{eqn:EoM} then yields predictions for the ballistic-scale time evolution of the true charge density given by $\langle \hat{n}(\mathbf{x},t)\rangle$.

\section{Numerical methods}
\label{appendix:numericalmethods}
\subsection{Staggered leapfrog method}
A real-space numerical scheme that we have found suitable for integrating both Eqs.~\eqref{eqn:EoM} and its time-dependent generalization in Eq.~\eqref{eqn:tdepWignerEoM} numerically is a modification of the standard second-order staggered leapfrog method~\cite{press2007numerical} to include the internal rotation term $-i[h,w]$, which prevents these equations from being in standard flux-conserving form. For simplicity we summarize the details in one spatial dimension, relevant to the plots shown in this paper. We first introduce spatial and temporal grids $x_j$ and $t_n$ with equal spacings $\Delta x$ and $\Delta t$ respectively. In the absence of an external potential, the numerical integration can be performed independently for each $k$ point and is therefore parallelizable. We found that $200$ $k$ points were sufficient to capture the initial charge and charge-current densities in this paper. We also set $\Delta x=1$ and varied $\Delta t$ depending on the difficulty of achieving convergence, which we defined for practical purposes as invariance under halving the time step. 

To present the scheme, we fix a $k$ point and use the shorthand $w^n_j = w(x_j,k,t_n)$ for the matrix-valued Wigner function at each spacetime grid point. After the specification of the initial condition $w(x,k,0)$, we obtain the spatial profile at time $t_1=\Delta t$ from a lax forward time-centered space (FTCS)~\cite{press2007numerical} update rule
\begin{align}
    &w^1_j = \frac{1}{2}(w^0_{j+1} + w^0_{j-1}) \nonumber \\ &- \frac{\Delta t}{4 \Delta x} \{\partial_k h(k,t), w^{0}_{j+1} -  w^0_{j-1}\} - i \Delta t [h(k,t),w_j^0].
\end{align}
The integration for all subsequent time steps $n>1$ is carried out using a second-order staggered-leapfrog-type update rule 
\begin{align}
    &w^{n+1}_j = w^{n-1}_j \nonumber \\ & - \frac{\Delta t}{2\Delta x}\{\partial_k h(k,t),w^n_{j+1} - w^n_{j-1}\} - 2 i \Delta t[h(k,t),w^n_j].
\end{align}

\label{appendix:leapfrog}
\subsection{GPU Fourier kernel}
\label{appendix:fourier}
One of the key computational advantages of the Wigner formalism lies in the forgotten information associated with the full density matrix representation. For a box of size $\mathbf{\Lambda} \equiv (\Lambda_{0},\Lambda_{1},\ldots,\Lambda_{d})$ and a uniform lattice grid $\mathbf{G} = \mathbf{Z}^{d}\cap\mathbf{\Lambda}$, the Wigner function as defined in Eq.~\eqref{eqn:OrbitalWignerFunction} $w_{\alpha\beta}(\bx,\bk)$ is represented by a tensor with $\mathbf{\Lambda}\cdot\mathbf{\Lambda} M^{2}$ entries ($M$ is the unit-cell dimension). Note, the same number of entries is required for a full density matrix description on the grid $\mathbf{G}$, but time evolution operates only on the unit cell vs on the entire density matrix. 

For translation invariant noninteracting systems, we can work in the Bloch Hamiltonian basis to effectively reduce the computations of the full density matrix. Indeed, even in the presence of a uniform electric field, the Bloch states are modified as in Appendix~\ref{appendix:extfield} and the savings are shared in both formalisms. However, for nonuniform electric fields (and more generally local potentials), the Wigner formalism can also incorporate them as forcing terms 
in a derivative expansion as in Eq.~\eqref{eqn:FieldEoM}. 

Regardless of method, Wigner or exact diagonalization (ED), the tensor structure lends itself to a GPU implementation~\cite{dataAvailability}. This is done as follows in PyTorch~\cite{paszke2017automatic} for the general time-dependent case:
\begin{enumerate}
    \item Express the Bloch Hamiltonian as a torch.stack of different time and momentum slices, i.e., a tensor representation of the Hamiltonian indexed by time and momentum.
    \item Initialize a tensor representation of the initial state using the approximate thermal state in Eq.~\eqref{eq:locthermwigner}.
    \item Fourier transform $\bx\rightarrow \bp$ via the built in torch.einsum function (allows for custom BZ boundary conditions)---the $\partial_{\bx}$ in Eq.~\eqref{eqn:full_eom} just pulls down an $i\bp$.
    \item Construct the Fourier kernel by solving the Wigner equation of motion \eqref{eqn:full_eom}. Practically, this is done by enlarging the unit cell such that (in 1D) $\bh(k) = e^{ik}\bh_{+} + e^{-ik}\bh_{-}+ \bh_{0}$. Then $\partial_{k}\bh(k) = -\partial_{k}^{3}\bh(k)$ and $\partial_{k}^{2}\bh(k) = -\partial_{k}^{4}\bh(k)$. This closes the sum in Eq.~\eqref{eqn:full_eom}, which permits a re-summation of even and odd terms
    \begin{align}
    &\mathds{E}_j(\bp,\bk)\!=\! \cos{\frac{p}{2}} \left[h_{j+}(k)+h_{j-}(k)\right]+\,h_{j0}\,\\
    &\mathds{O}_j(\bp,\bk)\!=\! \sin{\frac{p}{2}} \partial_{k}\left[h_{j+}(k)+h_{j-}(k)\right] \,
    \end{align}
    We can then solve the equation of motion (in the time-independent case) by 
    \begin{align}
        w(\bp,\bk,t) = e^{-i\bh(\bk + \bp/2)t}  w(\bp,\bk,0)\,e^{i\bh(\bk - \bp/2)t}\,.
    \end{align}
    Note: Rewriting the Hamiltonian in the above form allows us to compress the time evolution tensor from $\mathbf{\Lambda}\cdot\mathbf{\Lambda}M^{2}$ parameters to $2\vert\mathbf{\Lambda}\vert M^{2}$ parameters by interpolating the grids.
    
    \item In the case of a time dependent Hamiltonian $\bh(\bk\pm\bp/2, t)$, we Trotterize the evolution into steps \begin{align}
        U(t_{f},t_{0}) = \prod_{n = 0}^{T}e^{-\frac{i}{T}\bh[\bk\pm\bp/2, t_{0}+ (t_{f}-t_{0})n/T ]}\,.
    \end{align}
    The Trotterization can be instantiated as a torch.stack and grouped into pairs by the associative property of matrix multiplication, i.e. take $T = 2^{D}$ and pair $\bh(t_{i}), \bh(t_{i+1})$ such that 
    \begin{align}
        &\prod_{n = 0}^{T}\exp[-\frac{i}{T}\bh(\bk\pm\bp/2,t_{0}+ (t_{f}-t_{0})n/T )]\rightarrow\nonumber\\
        &\prod_{n = 0}^{T/2}\exp[-\frac{i}{T}\bh(\bk\pm\bp/2,t_{0}+ (t_{f}-t_{0})2n/T )] \nonumber \\& \,\,\,\,\,\times \exp[-\frac{i}{T}\bh(\bk\pm\bp/2,t_{0}+ (t_{f}-t_{0})(2n+1)/T )]\,,
    \end{align}
    repeating $D$ times, and the tower of tensors collapses to a single tensor over the grid $\mathbf{G}$ of $M\times M$ unitary operators. The exponential reduction in number of time slices is done in batches over the joint $\bp,\bk$ grid to save memory.
    \item The final state is inverse Fourier transformed back via the same torch.einsum operation.
\end{enumerate}
Note, the boundary conditions for Fourier kernel methods are notoriously difficult to match after numerical integration, e.g., Gibbs phenomena ~\cite{boyd2001chebyshev}. We avoid this by taking advantage of the enormous system sizes GPU computation permits us to access and simply truncate boundary terms. As a consequence, effects such as Thouless pumping that are sensitive to the boundary are verified using the real-space method of Appendix~\ref{appendix:leapfrog}, but all wave packet dynamics (e.g., in Fig.~\ref{fig:FlatBands}) can be computed far from the boundary.

\bibliography{main_arxiv_v2}

\end{document}